\documentclass[%
reprint,
superscriptaddress,
showpacs,
amsmath,amssymb,
aps,
prl,
longbibliography,
]{revtex4-1}

\usepackage{psfrag,graphicx,epsfig,color}
\usepackage{dcolumn}
\usepackage{bm}
\usepackage{natbib}
\usepackage{float}
\usepackage[usenames,dvipsnames,svgnames,table]{xcolor}
\usepackage{subfigure}
\usepackage{nicefrac}

\def\re    {R_\lambda}
\def\uu {{\mathbf{u}}}
\def\aa {{\mathbf{a}}}

\definecolor{mygreen}{rgb}{0,0.7,0.}

\begin{document}

\title{
Scaling of acceleration statistics
in high Reynolds number turbulence
}

\author{Dhawal Buaria }
\email[]{dhawal.buaria@nyu.edu}
\affiliation{Tandon School of Engineering, New York University, New York, NY 11201, USA}
\affiliation{Max Planck Institute for Dynamics and Self-Organization, 37077 G\"ottingen, Germany}
\author{Katepalli R. Sreenivasan}
\affiliation{Tandon School of Engineering, New York University, New York, NY 11201, USA}
\affiliation{Department of Physics and the Courant Institute of Mathematical Sciences, New York University,
New York, NY 10012, USA}

\date{\today}

\begin{abstract}

The scaling of acceleration statistics in turbulence
is examined by combining data from the literature with new data from well-resolved
direct numerical simulations of isotropic turbulence, significantly extending 
the Reynolds number range. The acceleration variance at higher Reynolds numbers 
departs from previous predictions based on multifractal models, which 
characterize Lagrangian intermittency as an extension of Eulerian intermittency. 
The disagreement is even more prominent for higher-order 
moments of the acceleration. Instead, starting from a known exact 
relation, we relate the scaling of acceleration variance to that of 
Eulerian fourth-order velocity gradient and velocity increment statistics. 
This prediction is in excellent agreement with the variance data. Our work
highlights the need for models that consider Lagrangian intermittency 
independent of the Eulerian counterpart. 

\end{abstract}

\maketitle


\paragraph*{Introduction:}
The acceleration of a fluid element in a turbulent flow, given by the Lagrangian derivative of the velocity, resulting from the balance of forces acting on it, is arguably the simplest descriptor of its motion. This is directly reflected in the Navier-Stokes equations:
\begin{align}
\aa = D\uu / Dt = - \nabla p + \nu \nabla^2 \uu  + \mathbf{f} \ ,
\label{eq:ns}
\end{align}
where, $\uu$ is the divergence-free velocity  ($\nabla \cdot \uu=0$), $p$ the 
kinematic pressure, $\nu$ is the kinematic viscosity 
and $\mathbf{f}$ is a
forcing-term. Besides its fundamental role in the
study of turbulence \cite{laPorta01, toschi:2009, stelzenmuller, buaria.rs},
understanding the statistics of acceleration is of paramount
importance for diverse range of applications constructed around stochastic modeling of 
transport phenomena in turbulence 
\cite{Sawford91, wyngaard, pope1994, wilson1996}. 
The application of Kolmogorov's 1941 phenomenology 
implies that the variance (and higher-order moments) of any 
acceleration component $a$ can be solely described by the 
mean-dissipation rate $\langle \epsilon \rangle$ and $\nu$ \cite{K41a,Heisenberg:48,Yaglom:49}:
\begin{align}
\langle a^2 \rangle =  
\frac{1}{3} \langle |\mathbf{a}|^2 \rangle =
a_0 \ \langle \epsilon \rangle^{3/2} \nu^{-1/2} \ ,
\end{align}
where $a_0$ is thought to be a universal constant.

However, extensive numerical and experimental work has shown that $a_0$ 
increases with Reynolds number
\cite{Yeung89, Vedula:99, gotoh99, Voth02, Sawford03, mordant2004, gylfason2004, 
yeung2006, Ishihara07}. Thus, obtaining data on $a_0$ and modeling 
its $\re$-variation has been a topic of considerable interest. 
While several theoretical works have focused
on acceleration statistics \cite{hill2002, reynolds2003, beck2007, bentkamp19}, 
the most notable procedure -- 
but whose validity should not be taken for granted -- 
stems from the multifractal model \cite{borgas93, chevillard2003, biferale2004}, 
which quantifies acceleration intermittency (and, in general, the intermittency 
of other Lagrangian quantities) by  adapting to the Lagrangian viewpoint the 
well-known Eulerian framework, based either on the energy dissipation rate \cite{Sreeni97} 
or velocity increments \cite{Frisch95}. A key result from this 
consideration is that $a_0 \sim \re^\chi$, $\chi \approx 0.135$, where $\re$
is the Taylor-scale Reynolds number.
While data from direct numerical simulations (DNS) and experiments do not 
directly display this power-law,
it was nevertheless presumed to 
to be asymptotically correct at very
large $\re$, and an empirical 
interpolation formula \cite{Sawford03, yeung2006},
\begin{align}
a_0 \simeq 
\frac{c_1 \re^{\chi} }{ (1 + c_2/\re^{1+\chi}) } \ , 
\label{eq:a2c}
\end{align}
with $\chi=0.135$, $c_1 = 1.9$, $c_2=85$, 
was suggested to fit the data,
showing reasonable success \cite{Sawford03}.
An alternative scaling: $a_0 \sim R_\lambda^{0.25}$ was proposed by Hill \cite{hill2002}, 
which was indistinguishable from Eq.\eqref{eq:a2c} at low-$\re$ \cite{Sawford03, Ishihara07}; 
we discuss the veracity of this proposal later.

\begin{figure}
\centering
\includegraphics[width=0.45\textwidth]{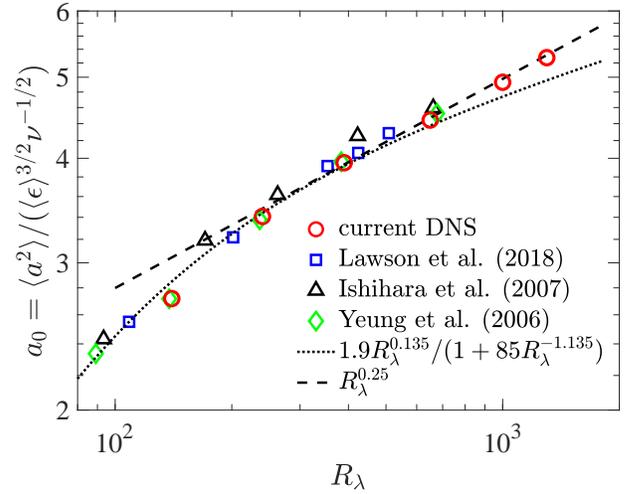} 
\caption{
Normalized acceleration 
variance $a_0 = \langle a^2\rangle / 
(\langle \epsilon \rangle^{3/2} \nu^{-1/2})$
as a function of $\re$. 
The data can be prescribed by a simple 
$\re^{0.25}$ power-law at high $\re$,
in contrast to the previously proposed 
empirical fit given by Eq.~\eqref{eq:a2c}.
}
\label{fig:acc2}
\end{figure}

In this Letter, we revisit the scaling of acceleration variance
(and higher-order moments) by 
presenting new DNS data at higher $\re$.
The new variance data agrees with previous lower $\re$ data, where the $\re$ range overlaps,
but increasing deviations from Eq.~\eqref{eq:a2c} occur at higher $\re$.
Results for high-order moments show even stronger deviations
from previous predictions.
Further analysis shows that 
the extension of Eulerian multifractal models
to the Lagrangian viewpoint
is the source of this discrepancy. 
We develop a statistical
model which shows excellent agreement
with variance data at high $\re$, and also
provide an updated interpolation fit 
to include low $\re$ data.

\paragraph*{Direct Numerical Simulations:}
The DNS data utilized here correspond to the canonical setup of forced stationary isotropic turbulence in a periodic domain \cite{Ishihara09}, 
allowing the use of highly accurate Fourier pseudo-spectral methods \cite{Rogallo}. 
The novelty is that we have simultaneously achieved very high Reynolds number and the necessary 
grid resolution to accurately resolve the small-scales \cite{BPBY2019,BS2020}. 
The data correspond to the same Taylor-scale Reynolds number $\re$ range of $140-1300$ 
attained in recent studies \cite{BBP2020, BPB2020, BP2021, BPB2022}, which have 
adequately established convergence 
with respect to resolution and statistical sampling. 
The grid resolution is as high as $k_{\rm max} \eta_K \approx 6$
which is substantially higher than 
$k_{\rm max} \eta_K \approx 1-2$, utilized in previous
acceleration studies \cite{Sawford03, yeung2006,Ishihara07};
$k_{\rm max} = \sqrt{2}N/3$ is the maximum resolved wavenumber on a $N^3$ grid
and $\eta_K= \nu^{3/4}\langle \epsilon \rangle^{-1/4} $ is the Kolmogorov length scale.
This improved small-scale resolution is 
especially necessary for capturing higher-order statistics of acceleration, since acceleration is even more intermittent than 
spatial velocity gradients \cite{Yeung89, yeung2006}.

\paragraph*{Acceleration variance:}Figure~\ref{fig:acc2} shows the compilation of data 
from various sources including data from both DNS \cite{yeung2006,Ishihara07}
\footnote{
Due to resolution concerns, we only use series 2
data from \cite{Ishihara07}, corresponding to 
$k_{\rm max}\eta_K \approx 2$}
and bias-corrected experiments \cite{lawson2018}.
We have also included DNS data obtained directly
from Lagrangian trajectories of fluid particles
\cite{BSY.2015, BYS.2016, buaria.cpc},
which give identical results for acceleration variance
\footnote{this data is at somewhat lower resolution
of $k_{\rm max}\eta_K\approx 1.5$, which does not 
effect the variance, but 
errors for higher-order moments are significant}.
As evident, while  Eq.~\eqref{eq:a2c} works for the previous range of $\re$, it does not fit the new data. In fact, a $\re^{0.25}$ scaling is more appropriate at higher $\re$, and as discussed later, 
the failure of multifractal models in fitting higher-order moments is even more conspicuous. 
To gain clarity on this point, it is useful to discuss the multifractal models first.

\paragraph*{Acceleration scaling from multifractals:}
The key idea in multifractal approaches is to quantify 
the intermittency of acceleration in terms of 
the intermittency of Eulerian velocity gradients or dissipation rate.
Assuming a simple phenomenological equivalence between
temporal and spatial derivatives, acceleration can be 
written in terms
of dissipation rate and viscosity as $a \sim \epsilon^{3/4} \nu^{-1/4}$.
Thus, the moments of acceleration are obtained as:
\begin{align}
\langle a^p \rangle \sim \langle \epsilon^{3p/4} \rangle \nu^{-p/4} \ .
\label{eq:acc_dis}
\end{align}
Alternatively, 
\begin{align}
\langle a^p \rangle / \langle a_{\rm K}^p \rangle \sim 
\langle \epsilon^{3p/4} \rangle / \langle \epsilon \rangle^{3p/4}  \ ,
\label{eq:acc_dis2}
\end{align}
where $a_{\rm K} = \langle \epsilon\rangle ^{3/4} \nu^{-1/4}$,
i.e., acceleration based on Kolmogorov variables.
Since Eulerian intermittency dictates
that $\langle \epsilon^q \rangle \ne \langle \epsilon \rangle^q$
for any $q\ne1$ \cite{Frisch95,Sreeni97},
the key assumption in its extension to Lagrangian
intermittency is that the $p$-th moment of 
acceleration scales as the
$(3p/4)$-th moment of $\epsilon$ \cite{borgas93, biferale2004}. 
The scaling of
$\langle \epsilon^q \rangle/ \langle \epsilon \rangle^q$
can be obtained by several approaches, 
all leading to similar results.
We briefly summarize a few approaches
below, with additional details
in the Supplementary Material \cite{supp}.

The most direct approach is to utilize
the multifractality of 
dissipation-rate \cite{MS91,borgas93}.
Within the multifractal framework, 
a scale-averaged dissipation $\epsilon_r$,
over scale $r$,
is assumed to be H\"older continuous: 
$\epsilon_r /\langle \epsilon \rangle \sim (r/L)^{\alpha-1}$,
where $\alpha$ is the local H\"older exponent, 
with a corresponding multifractal spectrum $F(\alpha)$
and $L$ is the large-scale length. 
Note, the 1D spectrum $f(\alpha)$ is more
common in the literature \cite{MS91}, 
which is simply: $f(\alpha) = F(\alpha)-2$.
Now, $\epsilon_r$ reduces to the true dissipation
for a viscous-cutoff defined as:
$r \simeq (\nu^3/\epsilon_r)^{1/4}$
or equivalently, $r/L \simeq Re^{-3/(3+\alpha)}$.
Here, $Re=u' L/\nu$,
$u'$ being the large-scale velocity; 
we also use $Re \sim \re^2$
and $\langle \epsilon \rangle \sim u'^3/L$ from dissipation
anomaly \cite{sreeni84}.

The above framework leads to the result:
\begin{align}
\langle \epsilon^q \rangle / 
\langle \epsilon \rangle^q \sim R_\lambda^{\tau_q} \ , \  \
\tau_q = \sup_\alpha \frac{6[q(1-\alpha) - 3 + F(\alpha)]}{3+\alpha} \ .
\label{eq:tauq}
\end{align}
An approximation for $F(\alpha)$,
such as the $p$-model \cite{MS87,MS91},
can be used to obtain $\tau_q$. 
The $p$-th moment of acceleration can then be
simply obtained as
\footnote{Note, this result corresponds to absolute moments, 
since we only considered the magnitude,
but the odd moments of acceleration components
are identically zero from symmetry.}
\begin{align}
\langle a^p \rangle / a_{\rm K}^p \sim R_\lambda^{\zeta_p} \ ,
\ \ \text{with} \ \ \zeta_p = \tau_{3p/4} \ .
\end{align}

\begin{table}
\begin{tabular}{l|c|c|c|c}
\hline    
$M_a$                                            & p-model & She-Leveque  & K62 log-normal &  DNS result \\
\hline
$\langle a^2 \rangle /a_{\rm K}^2$               & 0.135   & 0.140        & 0.140  & 0.25       \\
$\langle a^4 \rangle /a_{\rm K}^4$               & 0.943   & 1.00         & 1.13   & 1.60        \\
$\langle a^6 \rangle /a_{\rm K}^6$               & 2.06    & 2.30         & 2.95   & 3.95       \\
\hline
$\langle a^4 \rangle $/$\langle a^2 \rangle^2 $  & 0.673   & 0.720        & 0.850  & 1.10       \\
$\langle a^6 \rangle $/$\langle a^2 \rangle^3 $  & 1.66    & 1.88         & 2.53.  & 3.20        \\
\hline    
\end{tabular}
\caption{Scaling exponents $\zeta$ for 
$\re$-scaling of acceleration moments
$M_a \sim R_\lambda^\zeta$, as predicted 
from intermittency models, 
compared with current DNS results (see
Figs.~\ref{fig:acc2} and \ref{fig:acc46}).
}
\label{tab:ap}
\end{table}

Instead of dissipation, one can also start by
taking the velocity increment $\delta u_r$ over scale $r$
to be H\"older continuous:
$\delta u_r/u' \sim (r/L)^h$, 
where $h$ is the local H\"older exponent and 
$D(h)$ is the corresponding multifractal spectrum.
A scale-dependent 
dissipation rate $\epsilon_r$ can then be
defined as $\epsilon_r \sim (\delta u_r)^3/r$,
which reduces to the true dissipation
for the viscous-cutoff defined by 
the condition
$\delta u_r r /\nu \simeq 1$. 
This framework leads to the same
result as in Eq.~\eqref{eq:tauq},
corresponding to $\alpha=3h$
and $F(\alpha) = D(h)$. 
A well-known approximation for
$D(h)$ is given by the She-Leveque model \cite{SL94}. 
Finally, we can also use the Kolmogorov (1962) 
log-normal model  \cite{K62}, which gives 
$\tau_q = 3 \mu q(q-1)/4$,
even though it is untenable for very large $q$
\cite{Frisch95}.
Here, $\mu$ is the intermittency exponent, with experiments
and DNS suggesting $\mu\approx$ 0.25 \cite{sreeni93,BS2022}.

\begin{figure}
\begin{center}
\includegraphics[width=0.45\textwidth]{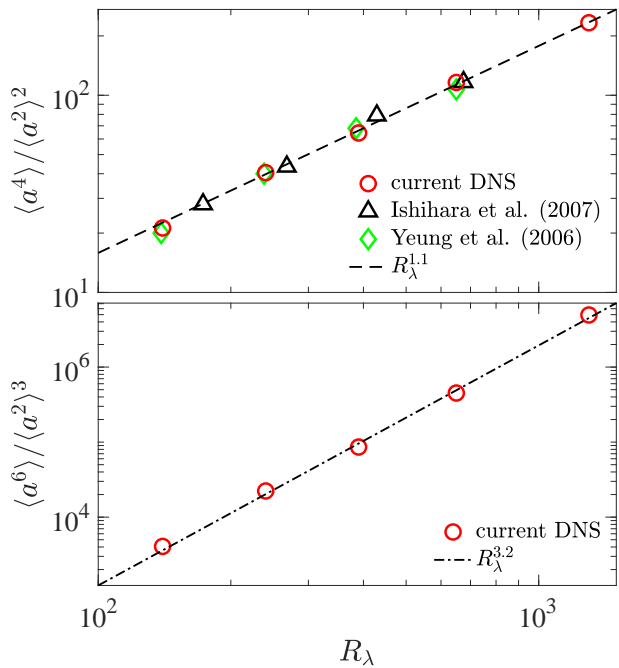}
\caption{
Normalized fourth (top) and sixth (bottom) order moments
of acceleration as a function of $\re$.
}
\label{fig:acc46}
\end{center}
\end{figure}

The scaling of acceleration moments obtained from these three approaches
and also from DNS data are listed in Table~\ref{tab:ap},
up to sixth-order. 
All approaches give essentially the same result for the
acceleration variance, with the exponent of about $0.135$ used in Eq.~\eqref{eq:a2c}.
However, the high-$\re$ DNS data clearly do not conform to any of the power laws shown in Table~\ref{tab:ap}.
The results for normalized fourth and sixth order moments, also plotted in Fig.~\ref{fig:acc46}, 
clearly show that the power-laws increasingly differ from multifractal predictions.

\begin{figure}
\begin{center}
\includegraphics[width=0.45\textwidth]{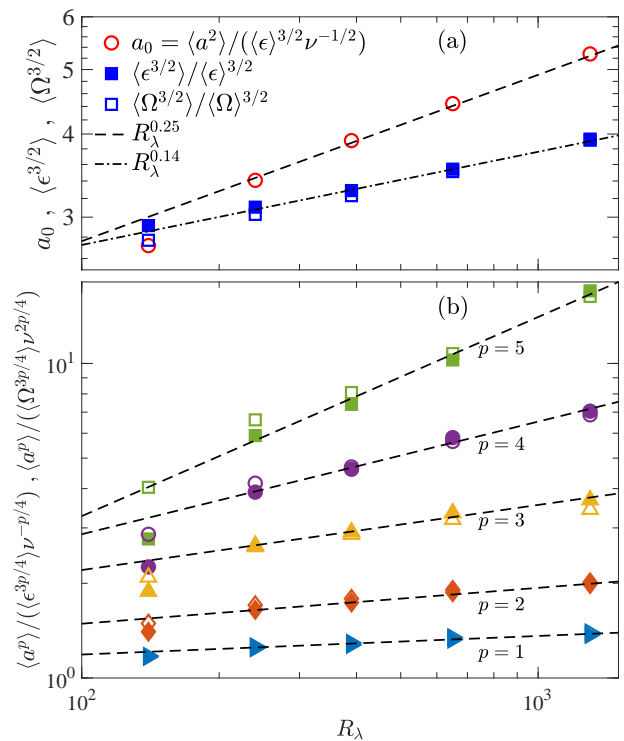} \\
\caption{
(a) Scaling of $a_0$,
$\langle \epsilon^{3/2} \rangle$ and $\langle \Omega^{3/2}\rangle$.
For clarity, data for $\epsilon$ and $\Omega$ are shifted up by a factor
of $2$ and $1.5$ respectively. 
(b) Scaling of $p$-th moments
of acceleration normalized by 
moments of $\epsilon^{3p/4}$ (filled symbols)
and $\Omega^{3p/4}$ (open symbols).
For clarity, 
data for $p=1$-$5$ are respectively shifted by factors
of 1, 0.75, 0.5, 0.25, 0.12 for $\epsilon$
and 0.94, 1.01, 1.2, 1.35, 1.6 for $\Omega$.
}
\label{fig:acc_dis}
\end{center}
\end{figure}

As noted earlier, the use of multifractals 
is primarily motivated by Eq.~\eqref{eq:acc_dis}.
To get better insight, in Fig.\ref{fig:acc_dis}a, we plot 
$a_0$ and 
$\langle \epsilon^{3/2} \rangle / \langle \epsilon \rangle^{3/2}$
versus $\re$. While the latter
shows a clear $\re^{0.14}$ scaling as anticipated from multifractals (and also the log-normal model), the 
former shows a steeper scaling of $R_\lambda^{0.25}$.
An even more general and direct test is presented in 
Fig.\ref{fig:acc_dis}b, by checking the validity of
Eq.~\eqref{eq:acc_dis} for different $p$ values.
The data clearly suggest that the acceleration intermittency, 
being stronger, cannot be described by 
extending the Eulerian intermittency
of the dissipation-rate. 
In fact, a similar observation
has been made for Lagrangian velocity
structure functions,
where extensions of the $p$-model
and the She-Leveque model severely underpredicts
their intermittency (i.e., overpredicts
the inertial-range exponents) \cite{sawford15}. 

It is worth considering if one might describe
the scaling of acceleration moments in terms of 
enstrophy $\Omega=|\boldsymbol{\omega}|^2$
($\boldsymbol{\omega}$ being the vorticity), instead
of dissipation.
This change addresses the likelihood that acceleration is influenced
more by transverse velocity gradients than by longitudinal ones \cite{arneodo,lanotte}.
In isotropic turbulence $\langle \Omega \rangle
= \langle \epsilon \rangle /\nu$, but the higher moments
differ; enstrophy being  more intermittent
\cite{BPBY2019, BP2022}. 
The resulting modification to Eq.~\eqref{eq:acc_dis}
is: $\langle a^p \rangle \sim \langle \Omega^{3p/4} \nu^{2p/4}$.
However, as tested in Fig.~\ref{fig:acc_dis}a,b, 
the differences
arise only for large $p$; even then, it is not
sufficient to explain the stronger intermittency
of acceleration (also see Supplementary \cite{supp}).

\paragraph*{Acceleration variance
from fourth-order structure function:}
A statistical model for acceleration variance
is now obtained using a methodology
similar to that proposed by \cite{hill2002}, but
differing in some crucial aspects.  
From Eq.~\eqref{eq:ns}, acceleration variance
can be obtained directly as
\footnote{The forcing term has negligible contribution 
to acceleration moments. This is 
also reaffirmed by collapse of data in Fig.~\ref{fig:acc2}
from various sources that use different forcings}
\begin{align}
\langle |\mathbf{a}|^2 \rangle = \langle |\nabla p|^2 \rangle
 + \nu^2 \langle | \nabla^2 u |^2 \rangle \ .
 \label{eq:a2ns}
\end{align}
The viscous contribution is known to be small
and can be ignored \cite{Vedula:99}.
An exact relation for variance of pressure-gradient
is also known \cite{MY.II, hill1995}:
\begin{align}
\langle |\nabla p|^2 \rangle = 
\int_r r^{-3} 
[ D_{1111}(r) + D_{\alpha \alpha \alpha \alpha} (r)
-6 D_{11\beta \beta}(r) ] dr \ ,
\label{eq:p2}
\end{align}
where the $D$s are the fourth order longitudinal, 
transverse and mixed structure functions, in order.
The above results can be rewritten as \cite{hill2002}:
\begin{align}
\langle |\mathbf{a}|^2 \rangle \simeq
4 H_\chi  \int_r  r^{-3} D_{1111} (r) dr,
\label{eq:d4int}
\end{align}
where $H_\chi$ is defined
by Eqs.~\eqref{eq:a2ns}-\eqref{eq:p2}.
At sufficiently high $\re$ ($\gtrsim200$), DNS data
\cite{Vedula:99, Ishihara07} confirm that
$H_\chi \approx 0.65$ (also see Supplementary \cite{supp}).
We can normalize both sides
by Kolmogorov-scales to write
\begin{align}
a_0 \simeq \frac{4 H_\chi}{3}  \int_r  \left( \frac{r}{\eta_K} \right)^{-3} 
\frac{D_{1111} (r)}{u_K^4} d \left( \frac{r}{\eta_K}\right) \ .
\label{eq:a0d4}
\end{align}
Assuming standard scaling regimes \cite{Frisch95},
we can write 
\begin{align}
\frac{D_{1111}(r)}{u_K^4} =
  \begin{cases}
\frac{F}{225} \left( \frac{r}{\eta_K} \right)^4 
\ \ \ \ \ \    &r < \ell \ , \\
C_4 \left( \frac{r}{\eta_K} \right)^{\xi_4} 
\ \ \ \ \ \    &\ell < r < L \ , \\
C 
\ \ \ \ \ \ \  &r > L \ , 
\end{cases}
\label{eq:d4}
\end{align}
where $F$ is the flatness of $\partial u/\partial x$, 
$\xi_4$ is the inertial-range
exponent, and $C_4$, $C$ are constants which depend on $\re$; 
$\ell$ is a crossover scale between the viscous and inertial range and
is determined by matching the two regimes as
\begin{align} 
(F/225) ( {\ell}/{\eta_K})^4 =
C_4 ( {\ell}/{\eta_K} )^{\xi_4} \ .
\end{align}

Now, taking
\begin{align}
F \sim \re^\alpha \ , \ \ \ \ \  C_4 \sim \re^\beta \ ,
\end{align}
we have
\begin{align} 
{\ell}/{\eta_K} \sim   \re^{(\beta - \alpha)/(4-\xi_4)} \ .
\end{align}
Finally, from piecewise integration 
of Eq.~\eqref{eq:a0d4},
it can be shown that
(see Supplementary Material \cite{supp} for 
intermediate steps):
\begin{align} 
a_0 \sim  F (\ell/\eta_K)^2
\end{align}
Substituting the $\re$-dependencies, we get
\begin{align} 
a_0 \sim  \re^{(2\alpha - \alpha \xi_4 + 2\beta)/(4-\xi_4)} \ .
\label{eq:a0f}
\end{align}

\begin{figure}
\begin{center}
\includegraphics[width=0.45\textwidth]{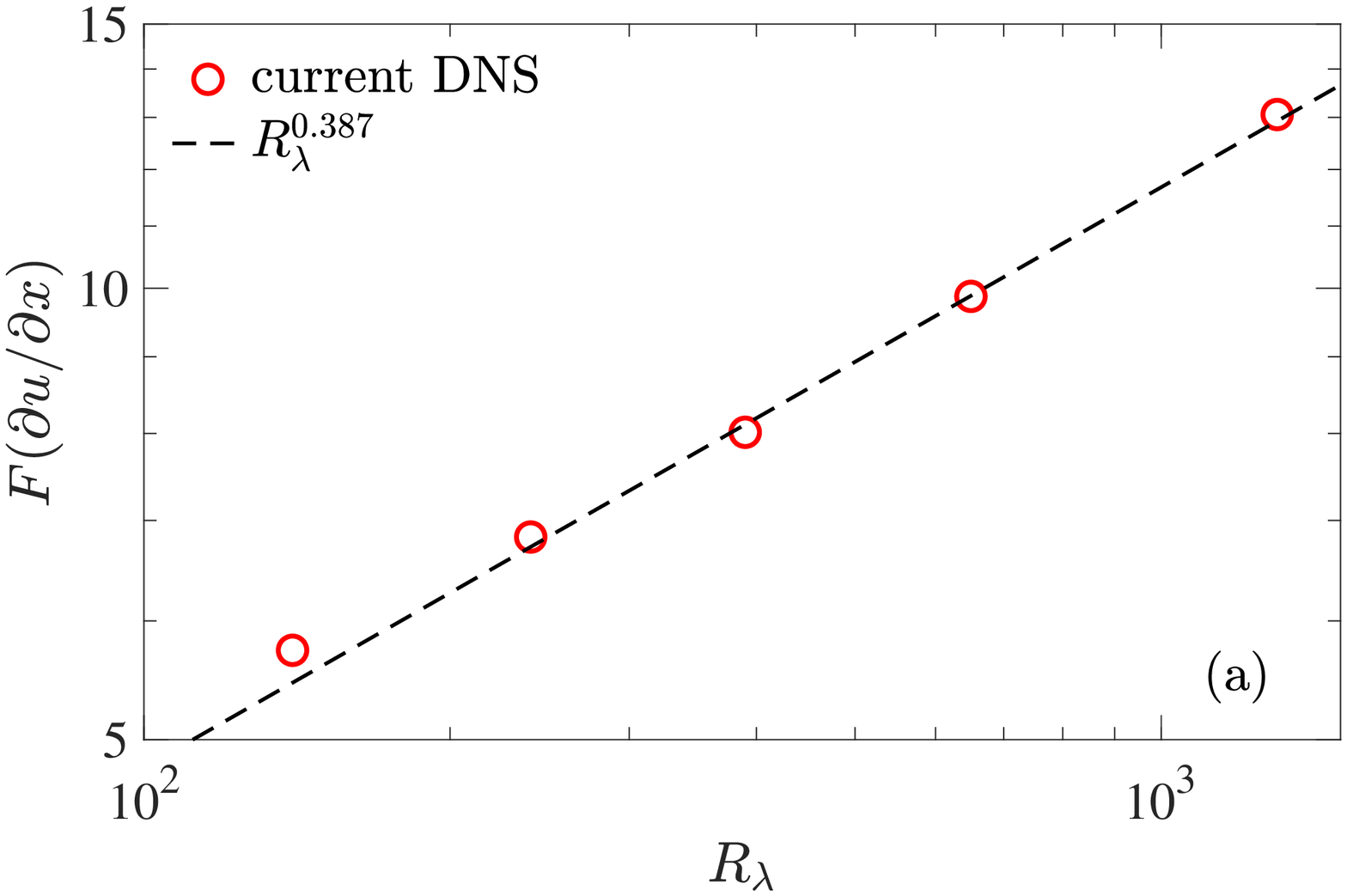} \\
\vspace{0.3cm}
\includegraphics[width=0.47\textwidth]{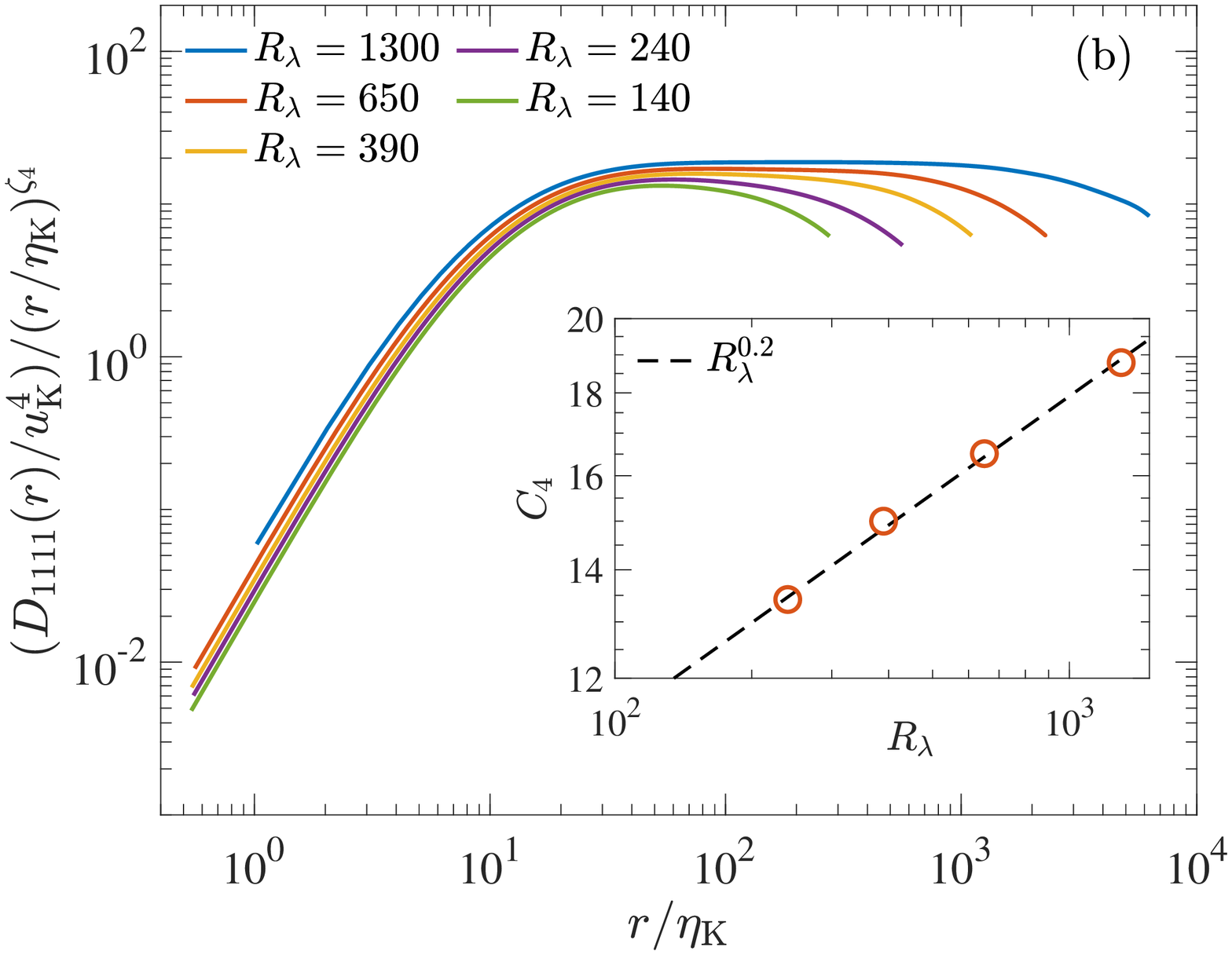}
\caption{
(a) Flatness of longitudinal velocity gradient
as a function of $\re$. 
(b) Fourth-order structure function compensated by its inertial-range scaling. The inset shows the variation of coefficient $C_4$ in Eq.~\eqref{eq:d4} as a function of $\re$. 
}
\label{fig:fdudx}
\end{center}
\end{figure}

The values of $\alpha$, $\beta$ and $\xi_4$ 
are in principle obtainable from Eulerian
intermittency models.
The exponent $\alpha$ simply corresponds
to $\tau_2$ in Eq.~\eqref{eq:tauq}, since
$F \sim \langle \epsilon^2 \rangle/\langle \epsilon \rangle^2$. 
Multifractal and log-normal models predict 
$\alpha = \tau_2 \approx 0.38$. 
The DNS data for $F$ are shown
in Fig.~\ref{fig:fdudx}a, giving 
$\alpha \approx 0.387$, in excellent
agreement with the prediction, and
also with previous 
experimental and DNS results
in literature \cite{gylfason2004,Ishihara07}.

On the other hand, intermittency
models predict $\xi_4\approx1.28$ \cite{SL94}.
Our DNS shows $\xi_4\approx 1.3$, which is
well within statistical error bounds.
Finally, the prediction for $\beta$
from multifractal model is
$\beta \approx (4- 3\xi_4)/2$, which reduces
to $\beta \approx \mu/3$ for log-normal model; both predictions
give $\beta \approx 0.08$ 
(also see Supplementary \cite{supp}). 
Figure~\ref{fig:fdudx}b shows
the normalized fourth-order structure function from 
our DNS data, using $\xi_4 \approx 1.3$. Note, as expected, the inertial-range increases 
with $\re$. The inset of the bottom panel shows $C_4$, giving $\beta \approx 0.2$. 
This observed $\beta$ is substantially larger than $0.08$ anticipated
from multifractal and log-normal models.

The use of
$\alpha=0.387$, $\beta=0.2$
and $\xi_4 = 1.3$ 
in Eq.~\eqref{eq:a0f} leads to
\begin{align} 
a_0 \sim  \re^{\chi}, \ \ \ \chi=\zeta_2 \approx 0.25 \ ,
\end{align}
which is in excellent agreement with the
high-$\re$ data shown in Figs.~\ref{fig:acc2} and \ref{fig:acc_dis}a.
The exponent $0.25$ is
virtually insensitive to a small
variation in $\xi_4$, but is significantly
impacted by the choice of $\beta=0.2$ 
(instead of $0.08$).
Moreover, the use of $\beta\approx0.08$ 
in Eq.~\eqref{eq:a0f} 
gives $a_0 \sim \re^{0.15}$, which
is essentially the same as the  
exponent $0.14$ obtained earlier in Table~\ref{tab:ap}. 
This shows the robustness of piecewise integration leading
to the result
in Eq.~\eqref{eq:a0f} and also suggests
that the discrepancy from multifractal
prediction is due to the exponent $\beta$
(and hence the proportionality constant $C_4$).
In this regard, the role of $\beta$ needs to be further
explored, especially
in relation to the inadequacy of Eq.~\eqref{eq:acc_dis}.

We note that the exponent $0.25$ was 
also suggested by Hill \cite{hill2002}. 
However, Hill arrived at this result 
by deriving that $a_0 \sim F^{0.79}$ and
$F\sim\re^{0.31}$ based on \cite{antonia1981};
evidently, the current data do not agree with both of these results. It appears that the two errors fortuitously cancelled out each other 
to give the $0.25$ exponent.
Finally, we point out that the exponent $0.25$ describes the data for $\re\gtrsim200$. 
To describe the data at lower $\re$, 
an empirical interpolation
formula in the spirit of Eq.~\eqref{eq:a2c} can be 
devised with $\chi=0.25$. Least-square fit
gives $c_1 \approx 0.89$, $c_2 \approx 40$
(also see Supplementary \cite{supp}).


\paragraph*{Conclusions:}
The moments of Lagrangian acceleration are known
to deviate from classical K41 phenomenology 
due to intermittency.
Attempts were made to quantify these deviations
by extending the Eulerian multifractal models 
to the Lagrangian viewpoint and devising 
an ad-hoc interpolation formula to agree with available
data from DNS and experiments.
The first contribution of this article is to present new 
very well resolved DNS data on Lagrangian acceleration at higher $\re$, and show that they disagree with the results from multifractal models, and the interpolation formula. The disagreement gets increasingly stronger with the moment
order. As a second contribution, the article devises a statistical 
model that is able to correctly capture the scaling of 
acceleration variance. 
While this framework does not seem amenable for generalization to higher-order moments, our
results show that the intermittency of Lagrangian
quantities remains an open problem, even more compellingly than before.

\begin{acknowledgements}
\paragraph*{Acknowledgments:}
We thank P.K. Yeung and Luca Biferale for providing
helpful comments on an earlier draft of the manuscript. 
We gratefully acknowledge the Gauss Centre for Supercomputing 
e.V. for providing computing time on the supercomputers 
JUQUEEN and JUWELS at J\"ulich Supercomputing Centre (JSC),
where the simulations reported in this
paper were performed.
\end{acknowledgements}


\begin{thebibliography}{60}%
\makeatletter
\providecommand \@ifxundefined [1]{%
 \@ifx{#1\undefined}
}%
\providecommand \@ifnum [1]{%
 \ifnum #1\expandafter \@firstoftwo
 \else \expandafter \@secondoftwo
 \fi
}%
\providecommand \@ifx [1]{%
 \ifx #1\expandafter \@firstoftwo
 \else \expandafter \@secondoftwo
 \fi
}%
\providecommand \natexlab [1]{#1}%
\providecommand \enquote  [1]{``#1''}%
\providecommand \bibnamefont  [1]{#1}%
\providecommand \bibfnamefont [1]{#1}%
\providecommand \citenamefont [1]{#1}%
\providecommand \href@noop [0]{\@secondoftwo}%
\providecommand \href [0]{\begingroup \@sanitize@url \@href}%
\providecommand \@href[1]{\@@startlink{#1}\@@href}%
\providecommand \@@href[1]{\endgroup#1\@@endlink}%
\providecommand \@sanitize@url [0]{\catcode `\\12\catcode `\$12\catcode
  `\&12\catcode `\#12\catcode `\^12\catcode `\_12\catcode `\%12\relax}%
\providecommand \@@startlink[1]{}%
\providecommand \@@endlink[0]{}%
\providecommand \url  [0]{\begingroup\@sanitize@url \@url }%
\providecommand \@url [1]{\endgroup\@href {#1}{\urlprefix }}%
\providecommand \urlprefix  [0]{URL }%
\providecommand \Eprint [0]{\href }%
\providecommand \doibase [0]{http://dx.doi.org/}%
\providecommand \selectlanguage [0]{\@gobble}%
\providecommand \bibinfo  [0]{\@secondoftwo}%
\providecommand \bibfield  [0]{\@secondoftwo}%
\providecommand \translation [1]{[#1]}%
\providecommand \BibitemOpen [0]{}%
\providecommand \bibitemStop [0]{}%
\providecommand \bibitemNoStop [0]{.\EOS\space}%
\providecommand \EOS [0]{\spacefactor3000\relax}%
\providecommand \BibitemShut  [1]{\csname bibitem#1\endcsname}%
\let\auto@bib@innerbib\@empty
\bibitem [{\citenamefont {La~Porta}\ \emph {et~al.}(2001)\citenamefont
  {La~Porta}, \citenamefont {Voth}, \citenamefont {Crawford}, \citenamefont
  {Alexander},\ and\ \citenamefont {Bodenschatz}}]{laPorta01}%
  \BibitemOpen
  \bibfield  {author} {\bibinfo {author} {\bibfnamefont {A.}~\bibnamefont
  {La~Porta}}, \bibinfo {author} {\bibfnamefont {G.~A.}\ \bibnamefont {Voth}},
  \bibinfo {author} {\bibfnamefont {A.~M.}\ \bibnamefont {Crawford}}, \bibinfo
  {author} {\bibfnamefont {J.}~\bibnamefont {Alexander}}, \ and\ \bibinfo
  {author} {\bibfnamefont {E.}~\bibnamefont {Bodenschatz}},\ }\bibfield
  {title} {\enquote {\bibinfo {title} {Fluid particle accelerations in fully
  developed turbulence},}\ }\href@noop {} {\bibfield  {journal} {\bibinfo
  {journal} {Nature}\ }\textbf {\bibinfo {volume} {409}},\ \bibinfo {pages}
  {1017--1019} (\bibinfo {year} {2001})}\BibitemShut {NoStop}%
\bibitem [{\citenamefont {Toschi}\ and\ \citenamefont
  {Bodenschatz}(2009)}]{toschi:2009}%
  \BibitemOpen
  \bibfield  {author} {\bibinfo {author} {\bibfnamefont {F.}~\bibnamefont
  {Toschi}}\ and\ \bibinfo {author} {\bibfnamefont {E.}~\bibnamefont
  {Bodenschatz}},\ }\bibfield  {title} {\enquote {\bibinfo {title}
  {{Lagrangian} properties of particles in turbulence},}\ }\href@noop {}
  {\bibfield  {journal} {\bibinfo  {journal} {Annu.~Rev.~Fluid Mech.}\ }\textbf
  {\bibinfo {volume} {41}},\ \bibinfo {pages} {375--404} (\bibinfo {year}
  {2009})}\BibitemShut {NoStop}%
\bibitem [{\citenamefont {Stelzenmuller}\ \emph {et~al.}(2017)\citenamefont
  {Stelzenmuller}, \citenamefont {Polanco}, \citenamefont {Vignal},
  \citenamefont {Vinkovic},\ and\ \citenamefont {Mordant}}]{stelzenmuller}%
  \BibitemOpen
  \bibfield  {author} {\bibinfo {author} {\bibfnamefont {N.}~\bibnamefont
  {Stelzenmuller}}, \bibinfo {author} {\bibfnamefont {J.~I.}\ \bibnamefont
  {Polanco}}, \bibinfo {author} {\bibfnamefont {L.}~\bibnamefont {Vignal}},
  \bibinfo {author} {\bibfnamefont {I.}~\bibnamefont {Vinkovic}}, \ and\
  \bibinfo {author} {\bibfnamefont {N.}~\bibnamefont {Mordant}},\ }\bibfield
  {title} {\enquote {\bibinfo {title} {Lagrangian acceleration statistics in a
  turbulent channel flow},}\ }\href@noop {} {\bibfield  {journal} {\bibinfo
  {journal} {Phys.~Rev.~Fluids}\ }\textbf {\bibinfo {volume} {2}},\ \bibinfo
  {pages} {054602} (\bibinfo {year} {2017})}\BibitemShut {NoStop}%
\bibitem [{\citenamefont {Buaria}\ \emph
  {et~al.}(2020{\natexlab{a}})\citenamefont {Buaria}, \citenamefont {Pumir},
  \citenamefont {Feraco}, \citenamefont {Marino}, \citenamefont {Pouquet},
  \citenamefont {Rosenberg},\ and\ \citenamefont {Primavera}}]{buaria.rs}%
  \BibitemOpen
  \bibfield  {author} {\bibinfo {author} {\bibfnamefont {D.}~\bibnamefont
  {Buaria}}, \bibinfo {author} {\bibfnamefont {A.}~\bibnamefont {Pumir}},
  \bibinfo {author} {\bibfnamefont {F.}~\bibnamefont {Feraco}}, \bibinfo
  {author} {\bibfnamefont {R.}~\bibnamefont {Marino}}, \bibinfo {author}
  {\bibfnamefont {A.}~\bibnamefont {Pouquet}}, \bibinfo {author} {\bibfnamefont
  {D.}~\bibnamefont {Rosenberg}}, \ and\ \bibinfo {author} {\bibfnamefont
  {L.}~\bibnamefont {Primavera}},\ }\bibfield  {title} {\enquote {\bibinfo
  {title} {Single-particle {Lagrangian} statistics from direct numerical
  simulations of rotating-stratified turbulence},}\ }\href@noop {} {\bibfield
  {journal} {\bibinfo  {journal} {Phys.~Rev.~Fluids}\ }\textbf {\bibinfo
  {volume} {5}},\ \bibinfo {pages} {064801} (\bibinfo {year}
  {2020}{\natexlab{a}})}\BibitemShut {NoStop}%
\bibitem [{\citenamefont {Sawford}(1991)}]{Sawford91}%
  \BibitemOpen
  \bibfield  {author} {\bibinfo {author} {\bibfnamefont {B.~L.}\ \bibnamefont
  {Sawford}},\ }\bibfield  {title} {\enquote {\bibinfo {title} {Reynolds number
  effects in {Lagrangian} stochastic models of turbulent dispersion},}\
  }\href@noop {} {\bibfield  {journal} {\bibinfo  {journal} {Phys. Fluids A}\
  }\textbf {\bibinfo {volume} {3}},\ \bibinfo {pages} {1577--1586} (\bibinfo
  {year} {1991})}\BibitemShut {NoStop}%
\bibitem [{\citenamefont {Wyngaard}(1992)}]{wyngaard}%
  \BibitemOpen
  \bibfield  {author} {\bibinfo {author} {\bibfnamefont {J.~C.}\ \bibnamefont
  {Wyngaard}},\ }\bibfield  {title} {\enquote {\bibinfo {title} {Atmospheric
  turbulence},}\ }\href@noop {} {\bibfield  {journal} {\bibinfo  {journal}
  {Annu.~Rev.~Fluid Mech.}\ }\textbf {\bibinfo {volume} {24}},\ \bibinfo
  {pages} {205--234} (\bibinfo {year} {1992})}\BibitemShut {NoStop}%
\bibitem [{\citenamefont {Pope}(1994)}]{pope1994}%
  \BibitemOpen
  \bibfield  {author} {\bibinfo {author} {\bibfnamefont {S.~B}\ \bibnamefont
  {Pope}},\ }\bibfield  {title} {\enquote {\bibinfo {title} {{Lagrangian} pdf
  methods for turbulent flows},}\ }\href@noop {} {\bibfield  {journal}
  {\bibinfo  {journal} {Annu.~Rev.~Fluid Mech.}\ }\textbf {\bibinfo {volume}
  {26}},\ \bibinfo {pages} {23--63} (\bibinfo {year} {1994})}\BibitemShut
  {NoStop}%
\bibitem [{\citenamefont {Wilson}\ and\ \citenamefont
  {Sawford}(1996)}]{wilson1996}%
  \BibitemOpen
  \bibfield  {author} {\bibinfo {author} {\bibfnamefont {J.~D.}\ \bibnamefont
  {Wilson}}\ and\ \bibinfo {author} {\bibfnamefont {B.~L.}\ \bibnamefont
  {Sawford}},\ }\href@noop {} {\enquote {\bibinfo {title} {Review of
  {Lagrangian} stochastic models for trajectories in the turbulent
  atmosphere},}\ } (\bibinfo {year} {1996})\BibitemShut {NoStop}%
\bibitem [{\citenamefont {Kolmogorov}(1941)}]{K41a}%
  \BibitemOpen
  \bibfield  {author} {\bibinfo {author} {\bibfnamefont {A.~N.}\ \bibnamefont
  {Kolmogorov}},\ }\bibfield  {title} {\enquote {\bibinfo {title} {Local
  structure of turbulence in an incompressible fluid for very large {Reynolds}
  numbers},}\ }\href@noop {} {\bibfield  {journal} {\bibinfo  {journal} {Dokl.
  Akad. Nauk. SSSR}\ }\textbf {\bibinfo {volume} {30}},\ \bibinfo {pages}
  {299--303} (\bibinfo {year} {1941})}\BibitemShut {NoStop}%
\bibitem [{\citenamefont {Heisenberg}(1948)}]{Heisenberg:48}%
  \BibitemOpen
  \bibfield  {author} {\bibinfo {author} {\bibfnamefont {W}~\bibnamefont
  {Heisenberg}},\ }\bibfield  {title} {\enquote {\bibinfo {title} {Zur
  statistichen {T}heorie der {T}urbulenz},}\ }\href@noop {} {\bibfield
  {journal} {\bibinfo  {journal} {Z Phys}\ }\textbf {\bibinfo {volume} {124}},\
  \bibinfo {pages} {628--657} (\bibinfo {year} {1948})}\BibitemShut {NoStop}%
\bibitem [{\citenamefont {Yaglom}(1949)}]{Yaglom:49}%
  \BibitemOpen
  \bibfield  {author} {\bibinfo {author} {\bibfnamefont {A.~M.}\ \bibnamefont
  {Yaglom}},\ }\bibfield  {title} {\enquote {\bibinfo {title} {On the
  acceleration field in a turbulent flow},}\ }\href@noop {} {\bibfield
  {journal} {\bibinfo  {journal} {C. R. Akad. Nauk. URSS}\ }\textbf {\bibinfo
  {volume} {67}},\ \bibinfo {pages} {795--798} (\bibinfo {year}
  {1949})}\BibitemShut {NoStop}%
\bibitem [{\citenamefont {Yeung}\ and\ \citenamefont {Pope}(1989)}]{Yeung89}%
  \BibitemOpen
  \bibfield  {author} {\bibinfo {author} {\bibfnamefont {P.~K.}\ \bibnamefont
  {Yeung}}\ and\ \bibinfo {author} {\bibfnamefont {S.~B.}\ \bibnamefont
  {Pope}},\ }\bibfield  {title} {\enquote {\bibinfo {title} {{Lagrangian}
  statistics from direct numerical simulations of isotropic turbulence},}\
  }\href@noop {} {\bibfield  {journal} {\bibinfo  {journal} {J. Fluid Mech.}\
  }\textbf {\bibinfo {volume} {207}},\ \bibinfo {pages} {531--586} (\bibinfo
  {year} {1989})}\BibitemShut {NoStop}%
\bibitem [{\citenamefont {Vedula}\ and\ \citenamefont
  {Yeung}(1999)}]{Vedula:99}%
  \BibitemOpen
  \bibfield  {author} {\bibinfo {author} {\bibfnamefont {P.}~\bibnamefont
  {Vedula}}\ and\ \bibinfo {author} {\bibfnamefont {P.~K.}\ \bibnamefont
  {Yeung}},\ }\bibfield  {title} {\enquote {\bibinfo {title} {Similarity
  scaling of acceleration and pressure statistics in numerical simulations of
  isotropic turbulence},}\ }\href@noop {} {\bibfield  {journal} {\bibinfo
  {journal} {Phys Fluids}\ }\textbf {\bibinfo {volume} {11}},\ \bibinfo {pages}
  {1208--1220} (\bibinfo {year} {1999})}\BibitemShut {NoStop}%
\bibitem [{\citenamefont {Gotoh}\ and\ \citenamefont
  {Rogallo}(1999)}]{gotoh99}%
  \BibitemOpen
  \bibfield  {author} {\bibinfo {author} {\bibfnamefont {T.}~\bibnamefont
  {Gotoh}}\ and\ \bibinfo {author} {\bibfnamefont {R.~S.}\ \bibnamefont
  {Rogallo}},\ }\bibfield  {title} {\enquote {\bibinfo {title} {Intermittency
  and scaling of pressure at small scales in forced isotropic turbulence},}\
  }\href@noop {} {\bibfield  {journal} {\bibinfo  {journal} {J.~Fluid Mech.}\
  }\textbf {\bibinfo {volume} {396}},\ \bibinfo {pages} {257--285} (\bibinfo
  {year} {1999})}\BibitemShut {NoStop}%
\bibitem [{\citenamefont {Voth}\ \emph {et~al.}(2002)\citenamefont {Voth},
  \citenamefont {La~Porta}, \citenamefont {Crawford}, \citenamefont
  {Alexander},\ and\ \citenamefont {Bodenschatz}}]{Voth02}%
  \BibitemOpen
  \bibfield  {author} {\bibinfo {author} {\bibfnamefont {G.~A.}\ \bibnamefont
  {Voth}}, \bibinfo {author} {\bibfnamefont {A.}~\bibnamefont {La~Porta}},
  \bibinfo {author} {\bibfnamefont {A.~M.}\ \bibnamefont {Crawford}}, \bibinfo
  {author} {\bibfnamefont {J.}~\bibnamefont {Alexander}}, \ and\ \bibinfo
  {author} {\bibfnamefont {E.}~\bibnamefont {Bodenschatz}},\ }\bibfield
  {title} {\enquote {\bibinfo {title} {Measurement of particle accelerations in
  fully developed turbulence},}\ }\href@noop {} {\bibfield  {journal} {\bibinfo
   {journal} {J.~Fluid Mech.}\ }\textbf {\bibinfo {volume} {469}},\ \bibinfo
  {pages} {121--160} (\bibinfo {year} {2002})}\BibitemShut {NoStop}%
\bibitem [{\citenamefont {Sawford}\ \emph {et~al.}(2003)\citenamefont
  {Sawford}, \citenamefont {Yeung}, \citenamefont {Borgas}, \citenamefont
  {Vedula}, \citenamefont {Porta}, \citenamefont {Crawford},\ and\
  \citenamefont {Bodenschatz}}]{Sawford03}%
  \BibitemOpen
  \bibfield  {author} {\bibinfo {author} {\bibfnamefont {B.~L.}\ \bibnamefont
  {Sawford}}, \bibinfo {author} {\bibfnamefont {P.~K.}\ \bibnamefont {Yeung}},
  \bibinfo {author} {\bibfnamefont {M.~S.}\ \bibnamefont {Borgas}}, \bibinfo
  {author} {\bibfnamefont {P.}~\bibnamefont {Vedula}}, \bibinfo {author}
  {\bibfnamefont {A.~La}\ \bibnamefont {Porta}}, \bibinfo {author}
  {\bibfnamefont {A.~M.}\ \bibnamefont {Crawford}}, \ and\ \bibinfo {author}
  {\bibfnamefont {E.}~\bibnamefont {Bodenschatz}},\ }\bibfield  {title}
  {\enquote {\bibinfo {title} {Conditional and unconditional acceleration
  statistics in turbulence},}\ }\href@noop {} {\bibfield  {journal} {\bibinfo
  {journal} {Phys. Fluids}\ }\textbf {\bibinfo {volume} {15}},\ \bibinfo
  {pages} {3478--3489} (\bibinfo {year} {2003})}\BibitemShut {NoStop}%
\bibitem [{\citenamefont {Mordant}\ \emph {et~al.}(2004)\citenamefont
  {Mordant}, \citenamefont {L{\'e}v{\^e}que},\ and\ \citenamefont
  {Pinton}}]{mordant2004}%
  \BibitemOpen
  \bibfield  {author} {\bibinfo {author} {\bibfnamefont {N.}~\bibnamefont
  {Mordant}}, \bibinfo {author} {\bibfnamefont {E.}~\bibnamefont
  {L{\'e}v{\^e}que}}, \ and\ \bibinfo {author} {\bibfnamefont {J.-F.}\
  \bibnamefont {Pinton}},\ }\bibfield  {title} {\enquote {\bibinfo {title}
  {Experimental and numerical study of the {Lagrangian} dynamics of high
  reynolds turbulence},}\ }\href@noop {} {\bibfield  {journal} {\bibinfo
  {journal} {New J.~Phys.}\ }\textbf {\bibinfo {volume} {6}},\ \bibinfo {pages}
  {116} (\bibinfo {year} {2004})}\BibitemShut {NoStop}%
\bibitem [{\citenamefont {Gylfason}\ \emph {et~al.}(2004)\citenamefont
  {Gylfason}, \citenamefont {Ayyalasomayajula},\ and\ \citenamefont
  {Warhaft}}]{gylfason2004}%
  \BibitemOpen
  \bibfield  {author} {\bibinfo {author} {\bibfnamefont {A.}~\bibnamefont
  {Gylfason}}, \bibinfo {author} {\bibfnamefont {S.}~\bibnamefont
  {Ayyalasomayajula}}, \ and\ \bibinfo {author} {\bibfnamefont
  {Z.}~\bibnamefont {Warhaft}},\ }\bibfield  {title} {\enquote {\bibinfo
  {title} {Intermittency, pressure and acceleration statistics from hot-wire
  measurements in wind-tunnel turbulence},}\ }\href@noop {} {\bibfield
  {journal} {\bibinfo  {journal} {J.~Fluid Mech.}\ }\textbf {\bibinfo {volume}
  {501}},\ \bibinfo {pages} {213--229} (\bibinfo {year} {2004})}\BibitemShut
  {NoStop}%
\bibitem [{\citenamefont {Yeung}\ \emph {et~al.}(2006)\citenamefont {Yeung},
  \citenamefont {Pope}, \citenamefont {Lamorgese},\ and\ \citenamefont
  {Donzis}}]{yeung2006}%
  \BibitemOpen
  \bibfield  {author} {\bibinfo {author} {\bibfnamefont {P.~K.}\ \bibnamefont
  {Yeung}}, \bibinfo {author} {\bibfnamefont {S.~B.}\ \bibnamefont {Pope}},
  \bibinfo {author} {\bibfnamefont {A.~G.}\ \bibnamefont {Lamorgese}}, \ and\
  \bibinfo {author} {\bibfnamefont {D.~A.}\ \bibnamefont {Donzis}},\ }\bibfield
   {title} {\enquote {\bibinfo {title} {Acceleration and dissipation statistics
  of numerically simulated isotropic turbulence},}\ }\href@noop {} {\bibfield
  {journal} {\bibinfo  {journal} {Physics of fluids}\ }\textbf {\bibinfo
  {volume} {18}},\ \bibinfo {pages} {065103} (\bibinfo {year}
  {2006})}\BibitemShut {NoStop}%
\bibitem [{\citenamefont {Ishihara}\ \emph {et~al.}(2007)\citenamefont
  {Ishihara}, \citenamefont {Kaneda}, \citenamefont {Yokokawa}, \citenamefont
  {Itakura},\ and\ \citenamefont {Uno}}]{Ishihara07}%
  \BibitemOpen
  \bibfield  {author} {\bibinfo {author} {\bibfnamefont {T.}~\bibnamefont
  {Ishihara}}, \bibinfo {author} {\bibfnamefont {Y.}~\bibnamefont {Kaneda}},
  \bibinfo {author} {\bibfnamefont {M.}~\bibnamefont {Yokokawa}}, \bibinfo
  {author} {\bibfnamefont {K.}~\bibnamefont {Itakura}}, \ and\ \bibinfo
  {author} {\bibfnamefont {A.}~\bibnamefont {Uno}},\ }\bibfield  {title}
  {\enquote {\bibinfo {title} {Small-scale statistics in high resolution of
  numerically isotropic turbulence},}\ }\href@noop {} {\bibfield  {journal}
  {\bibinfo  {journal} {J.~Fluid~Mech.}\ }\textbf {\bibinfo {volume} {592}},\
  \bibinfo {pages} {335--366} (\bibinfo {year} {2007})}\BibitemShut {NoStop}%
\bibitem [{\citenamefont {Hill}(2002)}]{hill2002}%
  \BibitemOpen
  \bibfield  {author} {\bibinfo {author} {\bibfnamefont {R.~J.}\ \bibnamefont
  {Hill}},\ }\bibfield  {title} {\enquote {\bibinfo {title} {Scaling of
  acceleration in locally isotropic turbulence},}\ }\href@noop {} {\bibfield
  {journal} {\bibinfo  {journal} {J.~Fluid Mech.}\ }\textbf {\bibinfo {volume}
  {452}},\ \bibinfo {pages} {361--370} (\bibinfo {year} {2002})}\BibitemShut
  {NoStop}%
\bibitem [{\citenamefont {Reynolds}(2003)}]{reynolds2003}%
  \BibitemOpen
  \bibfield  {author} {\bibinfo {author} {\bibfnamefont {A.~M.}\ \bibnamefont
  {Reynolds}},\ }\bibfield  {title} {\enquote {\bibinfo {title}
  {Superstatistical mechanics of tracer-particle motions in turbulence},}\
  }\href@noop {} {\bibfield  {journal} {\bibinfo  {journal} {Phys.~Rev.~Lett.}\
  }\textbf {\bibinfo {volume} {91}},\ \bibinfo {pages} {084503} (\bibinfo
  {year} {2003})}\BibitemShut {NoStop}%
\bibitem [{\citenamefont {Beck}(2007)}]{beck2007}%
  \BibitemOpen
  \bibfield  {author} {\bibinfo {author} {\bibfnamefont {C.}~\bibnamefont
  {Beck}},\ }\bibfield  {title} {\enquote {\bibinfo {title} {Statistics of
  three-dimensional {Lagrangian} turbulence},}\ }\href@noop {} {\bibfield
  {journal} {\bibinfo  {journal} {Phys.~Rev.~Lett.}\ }\textbf {\bibinfo
  {volume} {98}},\ \bibinfo {pages} {064502} (\bibinfo {year}
  {2007})}\BibitemShut {NoStop}%
\bibitem [{\citenamefont {Bentkamp}\ \emph {et~al.}(2019)\citenamefont
  {Bentkamp}, \citenamefont {Lalescu},\ and\ \citenamefont
  {Wilczek}}]{bentkamp19}%
  \BibitemOpen
  \bibfield  {author} {\bibinfo {author} {\bibfnamefont {L.}~\bibnamefont
  {Bentkamp}}, \bibinfo {author} {\bibfnamefont {C.~C.}\ \bibnamefont
  {Lalescu}}, \ and\ \bibinfo {author} {\bibfnamefont {M.}~\bibnamefont
  {Wilczek}},\ }\bibfield  {title} {\enquote {\bibinfo {title} {Persistent
  accelerations disentangle {Lagrangian} turbulence},}\ }\href@noop {}
  {\bibfield  {journal} {\bibinfo  {journal} {Nat. Commun.}\ }\textbf {\bibinfo
  {volume} {10}},\ \bibinfo {pages} {1--8} (\bibinfo {year}
  {2019})}\BibitemShut {NoStop}%
\bibitem [{\citenamefont {Borgas}(1993)}]{borgas93}%
  \BibitemOpen
  \bibfield  {author} {\bibinfo {author} {\bibfnamefont {M.~S.}\ \bibnamefont
  {Borgas}},\ }\bibfield  {title} {\enquote {\bibinfo {title} {The multifractal
  {Lagrangian} nature of turbulence},}\ }\href@noop {} {\bibfield  {journal}
  {\bibinfo  {journal} {Philos. Trans. R. Soc. A}\ }\textbf {\bibinfo {volume}
  {342}},\ \bibinfo {pages} {379--411} (\bibinfo {year} {1993})}\BibitemShut
  {NoStop}%
\bibitem [{\citenamefont {Chevillard}\ \emph {et~al.}(2003)\citenamefont
  {Chevillard}, \citenamefont {Roux}, \citenamefont {L{\'e}v{\^e}que},
  \citenamefont {Mordant}, \citenamefont {Pinton},\ and\ \citenamefont
  {Arn{\'e}odo}}]{chevillard2003}%
  \BibitemOpen
  \bibfield  {author} {\bibinfo {author} {\bibfnamefont {L.}~\bibnamefont
  {Chevillard}}, \bibinfo {author} {\bibfnamefont {S.~G.}\ \bibnamefont
  {Roux}}, \bibinfo {author} {\bibfnamefont {E.}~\bibnamefont
  {L{\'e}v{\^e}que}}, \bibinfo {author} {\bibfnamefont {N.}~\bibnamefont
  {Mordant}}, \bibinfo {author} {\bibfnamefont {J.-F.}\ \bibnamefont {Pinton}},
  \ and\ \bibinfo {author} {\bibfnamefont {A.}~\bibnamefont {Arn{\'e}odo}},\
  }\bibfield  {title} {\enquote {\bibinfo {title} {{Lagrangian} velocity
  statistics in turbulent flows: Effects of dissipation},}\ }\href@noop {}
  {\bibfield  {journal} {\bibinfo  {journal} {Phys.~Rev.~Lett.}\ }\textbf
  {\bibinfo {volume} {91}},\ \bibinfo {pages} {214502} (\bibinfo {year}
  {2003})}\BibitemShut {NoStop}%
\bibitem [{\citenamefont {Biferale}\ \emph {et~al.}(2004)\citenamefont
  {Biferale}, \citenamefont {Boffetta}, \citenamefont {Celani}, \citenamefont
  {Devenish}, \citenamefont {Lanotte},\ and\ \citenamefont
  {Toschi}}]{biferale2004}%
  \BibitemOpen
  \bibfield  {author} {\bibinfo {author} {\bibfnamefont {L.}~\bibnamefont
  {Biferale}}, \bibinfo {author} {\bibfnamefont {G.}~\bibnamefont {Boffetta}},
  \bibinfo {author} {\bibfnamefont {A.}~\bibnamefont {Celani}}, \bibinfo
  {author} {\bibfnamefont {B.~J.}\ \bibnamefont {Devenish}}, \bibinfo {author}
  {\bibfnamefont {A.}~\bibnamefont {Lanotte}}, \ and\ \bibinfo {author}
  {\bibfnamefont {F.}~\bibnamefont {Toschi}},\ }\bibfield  {title} {\enquote
  {\bibinfo {title} {Multifractal statistics of {Lagrangian} velocity and
  acceleration in turbulence},}\ }\href@noop {} {\bibfield  {journal} {\bibinfo
   {journal} {Phys.~Rev.~Lett.}\ }\textbf {\bibinfo {volume} {93}},\ \bibinfo
  {pages} {064502} (\bibinfo {year} {2004})}\BibitemShut {NoStop}%
\bibitem [{\citenamefont {Sreenivasan}\ and\ \citenamefont
  {Antonia}(1997)}]{Sreeni97}%
  \BibitemOpen
  \bibfield  {author} {\bibinfo {author} {\bibfnamefont {K.~R.}\ \bibnamefont
  {Sreenivasan}}\ and\ \bibinfo {author} {\bibfnamefont {R.~A.}\ \bibnamefont
  {Antonia}},\ }\bibfield  {title} {\enquote {\bibinfo {title} {The
  phenomenology of small-scale turbulence},}\ }\href@noop {} {\bibfield
  {journal} {\bibinfo  {journal} {Annu.~Rev.~Fluid~Mech.}\ }\textbf {\bibinfo
  {volume} {29}},\ \bibinfo {pages} {435--77} (\bibinfo {year}
  {1997})}\BibitemShut {NoStop}%
\bibitem [{\citenamefont {Frisch}(1995)}]{Frisch95}%
  \BibitemOpen
  \bibfield  {author} {\bibinfo {author} {\bibfnamefont {U.}~\bibnamefont
  {Frisch}},\ }\href@noop {} {\emph {\bibinfo {title} {Turbulence: the legacy
  of {Kolmogorov}}}}\ (\bibinfo  {publisher} {Cambridge University Press},\
  \bibinfo {address} {Cambridge},\ \bibinfo {year} {1995})\BibitemShut
  {NoStop}%
\bibitem [{\citenamefont {Ishihara}\ \emph {et~al.}(2009)\citenamefont
  {Ishihara}, \citenamefont {Gotoh},\ and\ \citenamefont
  {Kaneda}}]{Ishihara09}%
  \BibitemOpen
  \bibfield  {author} {\bibinfo {author} {\bibfnamefont {T.}~\bibnamefont
  {Ishihara}}, \bibinfo {author} {\bibfnamefont {T.}~\bibnamefont {Gotoh}}, \
  and\ \bibinfo {author} {\bibfnamefont {Y.}~\bibnamefont {Kaneda}},\
  }\bibfield  {title} {\enquote {\bibinfo {title} {Study of high-{Reynolds}
  number isotropic turbulence by direct numerical simulations},}\ }\href@noop
  {} {\bibfield  {journal} {\bibinfo  {journal} {Ann.~Rev.~Fluid~Mech.}\
  }\textbf {\bibinfo {volume} {41}},\ \bibinfo {pages} {165--80} (\bibinfo
  {year} {2009})}\BibitemShut {NoStop}%
\bibitem [{\citenamefont {Rogallo}(1981)}]{Rogallo}%
  \BibitemOpen
  \bibfield  {author} {\bibinfo {author} {\bibfnamefont {R.~S.}\ \bibnamefont
  {Rogallo}},\ }\bibfield  {title} {\enquote {\bibinfo {title} {Numerical
  experiments in homogeneous turbulence},}\ }\href@noop {} {\bibfield
  {journal} {\bibinfo  {journal} {NASA Technical Memo}\ } (\bibinfo {year}
  {1981})}\BibitemShut {NoStop}%
\bibitem [{\citenamefont {Buaria}\ \emph {et~al.}(2019)\citenamefont {Buaria},
  \citenamefont {Pumir}, \citenamefont {Bodenschatz},\ and\ \citenamefont
  {Yeung}}]{BPBY2019}%
  \BibitemOpen
  \bibfield  {author} {\bibinfo {author} {\bibfnamefont {D.}~\bibnamefont
  {Buaria}}, \bibinfo {author} {\bibfnamefont {A.}~\bibnamefont {Pumir}},
  \bibinfo {author} {\bibfnamefont {E.}~\bibnamefont {Bodenschatz}}, \ and\
  \bibinfo {author} {\bibfnamefont {P.~K.}\ \bibnamefont {Yeung}},\ }\bibfield
  {title} {\enquote {\bibinfo {title} {Extreme velocity gradients in turbulent
  flows},}\ }\href@noop {} {\bibfield  {journal} {\bibinfo  {journal} {New
  J.~Phys.}\ }\textbf {\bibinfo {volume} {21}},\ \bibinfo {pages} {043004}
  (\bibinfo {year} {2019})}\BibitemShut {NoStop}%
\bibitem [{\citenamefont {Buaria}\ and\ \citenamefont
  {Sreenivasan}(2020)}]{BS2020}%
  \BibitemOpen
  \bibfield  {author} {\bibinfo {author} {\bibfnamefont {D.}~\bibnamefont
  {Buaria}}\ and\ \bibinfo {author} {\bibfnamefont {K.~R.}\ \bibnamefont
  {Sreenivasan}},\ }\bibfield  {title} {\enquote {\bibinfo {title} {Dissipation
  range of the energy spectrum in high {Reynolds} number turbulence},}\
  }\href@noop {} {\bibfield  {journal} {\bibinfo  {journal}
  {Phys.~Rev.~Fluids}\ }\textbf {\bibinfo {volume} {5}},\ \bibinfo {pages}
  {092601(R)} (\bibinfo {year} {2020})}\BibitemShut {NoStop}%
\bibitem [{\citenamefont {Buaria}\ \emph
  {et~al.}(2020{\natexlab{b}})\citenamefont {Buaria}, \citenamefont
  {Bodenschatz},\ and\ \citenamefont {Pumir}}]{BBP2020}%
  \BibitemOpen
  \bibfield  {author} {\bibinfo {author} {\bibfnamefont {D.}~\bibnamefont
  {Buaria}}, \bibinfo {author} {\bibfnamefont {E.}~\bibnamefont {Bodenschatz}},
  \ and\ \bibinfo {author} {\bibfnamefont {A.}~\bibnamefont {Pumir}},\
  }\bibfield  {title} {\enquote {\bibinfo {title} {Vortex stretching and
  enstrophy production in high {Reynolds} number turbulence},}\ }\href@noop {}
  {\bibfield  {journal} {\bibinfo  {journal} {Phys.~Rev.~Fluids}\ }\textbf
  {\bibinfo {volume} {5}},\ \bibinfo {pages} {104602} (\bibinfo {year}
  {2020}{\natexlab{b}})}\BibitemShut {NoStop}%
\bibitem [{\citenamefont {Buaria}\ \emph
  {et~al.}(2020{\natexlab{c}})\citenamefont {Buaria}, \citenamefont {Pumir},\
  and\ \citenamefont {Bodenschatz}}]{BPB2020}%
  \BibitemOpen
  \bibfield  {author} {\bibinfo {author} {\bibfnamefont {D.}~\bibnamefont
  {Buaria}}, \bibinfo {author} {\bibfnamefont {A.}~\bibnamefont {Pumir}}, \
  and\ \bibinfo {author} {\bibfnamefont {E.}~\bibnamefont {Bodenschatz}},\
  }\bibfield  {title} {\enquote {\bibinfo {title} {Self-attenuation of extreme
  events in {Navier-Stokes} turbulence},}\ }\href@noop {} {\bibfield  {journal}
  {\bibinfo  {journal} {Nat. Commun.}\ }\textbf {\bibinfo {volume} {11}},\
  \bibinfo {pages} {5852} (\bibinfo {year} {2020}{\natexlab{c}})}\BibitemShut
  {NoStop}%
\bibitem [{\citenamefont {Buaria}\ and\ \citenamefont {Pumir}(2021)}]{BP2021}%
  \BibitemOpen
  \bibfield  {author} {\bibinfo {author} {\bibfnamefont {D.}~\bibnamefont
  {Buaria}}\ and\ \bibinfo {author} {\bibfnamefont {A.}~\bibnamefont {Pumir}},\
  }\bibfield  {title} {\enquote {\bibinfo {title} {Nonlocal amplification of
  intense vorticity in turbulent flows},}\ }\href@noop {} {\bibfield  {journal}
  {\bibinfo  {journal} {Phys.~Rev.~Research}\ }\textbf {\bibinfo {volume}
  {3}},\ \bibinfo {pages} {042020} (\bibinfo {year} {2021})}\BibitemShut
  {NoStop}%
\bibitem [{\citenamefont {Buaria}\ \emph {et~al.}(2022)\citenamefont {Buaria},
  \citenamefont {Pumir},\ and\ \citenamefont {Bodenschatz}}]{BPB2022}%
  \BibitemOpen
  \bibfield  {author} {\bibinfo {author} {\bibfnamefont {D.}~\bibnamefont
  {Buaria}}, \bibinfo {author} {\bibfnamefont {A.}~\bibnamefont {Pumir}}, \
  and\ \bibinfo {author} {\bibfnamefont {E.}~\bibnamefont {Bodenschatz}},\
  }\bibfield  {title} {\enquote {\bibinfo {title} {Generation of intense
  dissipation in high {Reynolds} number turbulence},}\ }\href@noop {}
  {\bibfield  {journal} {\bibinfo  {journal} {Philos. Trans. R. Soc. A}\
  }\textbf {\bibinfo {volume} {380}},\ \bibinfo {pages} {20210088} (\bibinfo
  {year} {2022})}\BibitemShut {NoStop}%
\bibitem [{Note1()}]{Note1}%
  \BibitemOpen
  \bibinfo {note} {Due to resolution concerns, we only use series 2 data from
  \cite {Ishihara07}, corresponding to $k_{\protect \rm max}\eta _K \approx
  2$}\BibitemShut {NoStop}%
\bibitem [{\citenamefont {Lawson}\ \emph {et~al.}(2018)\citenamefont {Lawson},
  \citenamefont {Bodenschatz}, \citenamefont {Lalescu},\ and\ \citenamefont
  {Wilczek}}]{lawson2018}%
  \BibitemOpen
  \bibfield  {author} {\bibinfo {author} {\bibfnamefont {J.~M.}\ \bibnamefont
  {Lawson}}, \bibinfo {author} {\bibfnamefont {E.}~\bibnamefont {Bodenschatz}},
  \bibinfo {author} {\bibfnamefont {C.~C.}\ \bibnamefont {Lalescu}}, \ and\
  \bibinfo {author} {\bibfnamefont {M.}~\bibnamefont {Wilczek}},\ }\bibfield
  {title} {\enquote {\bibinfo {title} {Bias in particle tracking acceleration
  measurement},}\ }\href@noop {} {\bibfield  {journal} {\bibinfo  {journal}
  {Experiments in Fluids}\ }\textbf {\bibinfo {volume} {59}},\ \bibinfo {pages}
  {1--14} (\bibinfo {year} {2018})}\BibitemShut {NoStop}%
\bibitem [{\citenamefont {Buaria}\ \emph {et~al.}(2015)\citenamefont {Buaria},
  \citenamefont {Sawford},\ and\ \citenamefont {Yeung}}]{BSY.2015}%
  \BibitemOpen
  \bibfield  {author} {\bibinfo {author} {\bibfnamefont {D.}~\bibnamefont
  {Buaria}}, \bibinfo {author} {\bibfnamefont {B.~L.}\ \bibnamefont {Sawford}},
  \ and\ \bibinfo {author} {\bibfnamefont {P.~K.}\ \bibnamefont {Yeung}},\
  }\bibfield  {title} {\enquote {\bibinfo {title} {Characteristics of backward
  and forward two-particle relative dispersion in turbulence at different
  {R}eynolds numbers},}\ }\href@noop {} {\bibfield  {journal} {\bibinfo
  {journal} {Phys. Fluids}\ }\textbf {\bibinfo {volume} {27}},\ \bibinfo
  {pages} {105101} (\bibinfo {year} {2015})}\BibitemShut {NoStop}%
\bibitem [{\citenamefont {Buaria}\ \emph {et~al.}(2016)\citenamefont {Buaria},
  \citenamefont {Yeung},\ and\ \citenamefont {Sawford}}]{BYS.2016}%
  \BibitemOpen
  \bibfield  {author} {\bibinfo {author} {\bibfnamefont {D.}~\bibnamefont
  {Buaria}}, \bibinfo {author} {\bibfnamefont {P.~K.}\ \bibnamefont {Yeung}}, \
  and\ \bibinfo {author} {\bibfnamefont {B.~L.}\ \bibnamefont {Sawford}},\
  }\bibfield  {title} {\enquote {\bibinfo {title} {{A Lagrangian} study of
  turbulent mixing: forward and backward dispersion of molecular trajectories
  in isotropic turbulence},}\ }\href@noop {} {\bibfield  {journal} {\bibinfo
  {journal} {J.~Fluid Mech.}\ }\textbf {\bibinfo {volume} {{799}}},\ \bibinfo
  {pages} {{352--382}} (\bibinfo {year} {2016})}\BibitemShut {NoStop}%
\bibitem [{\citenamefont {Buaria}\ and\ \citenamefont
  {Yeung}(2017)}]{buaria.cpc}%
  \BibitemOpen
  \bibfield  {author} {\bibinfo {author} {\bibfnamefont {D.}~\bibnamefont
  {Buaria}}\ and\ \bibinfo {author} {\bibfnamefont {P.~K.}\ \bibnamefont
  {Yeung}},\ }\bibfield  {title} {\enquote {\bibinfo {title} {A highly scalable
  particle tracking algorithm using partitioned global address space ({PGAS})
  programming for extreme-scale turbulence simulations},}\ }\href@noop {}
  {\bibfield  {journal} {\bibinfo  {journal} {Comput. Phys. Commun.}\ }\textbf
  {\bibinfo {volume} {221}},\ \bibinfo {pages} {246 -- 258} (\bibinfo {year}
  {2017})}\BibitemShut {NoStop}%
\bibitem [{Note2()}]{Note2}%
  \BibitemOpen
  \bibinfo {note} {This data is at somewhat lower resolution of $k_{\protect
  \rm max}\eta _K\approx 1.5$, which does not effect the variance, but errors
  for higher-order moments are significant}\BibitemShut {NoStop}%
\bibitem [{sup()}]{supp}%
  \BibitemOpen
  \href@noop {} {\enquote {\bibinfo {title} {see {S}upplementary material for
  additional details},}\ }\BibitemShut {NoStop}%
\bibitem [{\citenamefont {Meneveau}\ and\ \citenamefont
  {Sreenivasan}(1991)}]{MS91}%
  \BibitemOpen
  \bibfield  {author} {\bibinfo {author} {\bibfnamefont {C.}~\bibnamefont
  {Meneveau}}\ and\ \bibinfo {author} {\bibfnamefont {K.~R.}\ \bibnamefont
  {Sreenivasan}},\ }\bibfield  {title} {\enquote {\bibinfo {title} {The
  multifractal nature of turbulent energy dissipation},}\ }\href@noop {}
  {\bibfield  {journal} {\bibinfo  {journal} {J.~Fluid Mech.}\ }\textbf
  {\bibinfo {volume} {224}},\ \bibinfo {pages} {429–--484} (\bibinfo {year}
  {1991})}\BibitemShut {NoStop}%
\bibitem [{\citenamefont {Sreenivasan}(1984)}]{sreeni84}%
  \BibitemOpen
  \bibfield  {author} {\bibinfo {author} {\bibfnamefont {K.~R.}\ \bibnamefont
  {Sreenivasan}},\ }\bibfield  {title} {\enquote {\bibinfo {title} {On the
  scaling of the turbulence energy dissipation rate},}\ }\href@noop {}
  {\bibfield  {journal} {\bibinfo  {journal} {Phys.~Fluids}\ }\textbf {\bibinfo
  {volume} {27}},\ \bibinfo {pages} {1048--1051} (\bibinfo {year}
  {1984})}\BibitemShut {NoStop}%
\bibitem [{\citenamefont {Meneveau}\ and\ \citenamefont
  {Sreenivasan}(1987)}]{MS87}%
  \BibitemOpen
  \bibfield  {author} {\bibinfo {author} {\bibfnamefont {C.}~\bibnamefont
  {Meneveau}}\ and\ \bibinfo {author} {\bibfnamefont {K.~R.}\ \bibnamefont
  {Sreenivasan}},\ }\bibfield  {title} {\enquote {\bibinfo {title} {Simple
  multifractal cascade model for fully developed turbulence},}\ }\href@noop {}
  {\bibfield  {journal} {\bibinfo  {journal} {Phys.~Rev.~Lett.}\ }\textbf
  {\bibinfo {volume} {59}},\ \bibinfo {pages} {1424} (\bibinfo {year}
  {1987})}\BibitemShut {NoStop}%
\bibitem [{Note3()}]{Note3}%
  \BibitemOpen
  \bibinfo {note} {Note, this result corresponds to absolute moments, since we
  only considered the magnitude, but the odd moments of acceleration components
  are identically zero from symmetry.}\BibitemShut {Stop}%
\bibitem [{\citenamefont {She}\ and\ \citenamefont {Leveque}(1994)}]{SL94}%
  \BibitemOpen
  \bibfield  {author} {\bibinfo {author} {\bibfnamefont {Z.-S.}\ \bibnamefont
  {She}}\ and\ \bibinfo {author} {\bibfnamefont {E.}~\bibnamefont {Leveque}},\
  }\bibfield  {title} {\enquote {\bibinfo {title} {Universal scaling laws in
  fully developed turbulence},}\ }\href@noop {} {\bibfield  {journal} {\bibinfo
   {journal} {Phys. Rev. Lett.}\ }\textbf {\bibinfo {volume} {72}},\ \bibinfo
  {pages} {336--339} (\bibinfo {year} {1994})}\BibitemShut {NoStop}%
\bibitem [{\citenamefont {Kolmogorov}(1962)}]{K62}%
  \BibitemOpen
  \bibfield  {author} {\bibinfo {author} {\bibfnamefont {A.~N.}\ \bibnamefont
  {Kolmogorov}},\ }\bibfield  {title} {\enquote {\bibinfo {title} {A refinement
  of previous hypotheses concerning the local structure of turbulence in a
  viscous incompressible fluid at high {Reynolds} number},}\ }\href@noop {}
  {\bibfield  {journal} {\bibinfo  {journal} {J.~Fluid Mech.}\ }\textbf
  {\bibinfo {volume} {13}},\ \bibinfo {pages} {82--85} (\bibinfo {year}
  {1962})}\BibitemShut {NoStop}%
\bibitem [{\citenamefont {Sreenivasan}\ and\ \citenamefont
  {Kailasnath}(1993)}]{sreeni93}%
  \BibitemOpen
  \bibfield  {author} {\bibinfo {author} {\bibfnamefont {K.~R.}\ \bibnamefont
  {Sreenivasan}}\ and\ \bibinfo {author} {\bibfnamefont {P.}~\bibnamefont
  {Kailasnath}},\ }\bibfield  {title} {\enquote {\bibinfo {title} {An update on
  the intermittency exponent in turbulence},}\ }\href@noop {} {\bibfield
  {journal} {\bibinfo  {journal} {Phys. Fluids A: Fluid Dynamics}\ }\textbf
  {\bibinfo {volume} {5}},\ \bibinfo {pages} {512--514} (\bibinfo {year}
  {1993})}\BibitemShut {NoStop}%
\bibitem [{\citenamefont {Buaria}\ and\ \citenamefont
  {Sreenivasan}(2022)}]{BS2022}%
  \BibitemOpen
  \bibfield  {author} {\bibinfo {author} {\bibfnamefont {D.}~\bibnamefont
  {Buaria}}\ and\ \bibinfo {author} {\bibfnamefont {K.~R.}\ \bibnamefont
  {Sreenivasan}},\ }\bibfield  {title} {\enquote {\bibinfo {title}
  {Intermittency of turbulent velocity and scalar fields using {3D} local
  averaging},}\ }\href@noop {} {\bibfield  {journal} {\bibinfo  {journal}
  {arXiv:2204.08132}\ } (\bibinfo {year} {2022})}\BibitemShut {NoStop}%
\bibitem [{\citenamefont {Sawford}\ and\ \citenamefont
  {Yeung}(2015)}]{sawford15}%
  \BibitemOpen
  \bibfield  {author} {\bibinfo {author} {\bibfnamefont {B.~L.}\ \bibnamefont
  {Sawford}}\ and\ \bibinfo {author} {\bibfnamefont {P.~K.}\ \bibnamefont
  {Yeung}},\ }\bibfield  {title} {\enquote {\bibinfo {title} {Direct numerical
  simulation studies of {Lagrangian} intermittency in turbulence},}\
  }\href@noop {} {\bibfield  {journal} {\bibinfo  {journal} {Phys.~Fluids}\
  }\textbf {\bibinfo {volume} {27}},\ \bibinfo {pages} {065109} (\bibinfo
  {year} {2015})}\BibitemShut {NoStop}%
\bibitem [{\citenamefont {Arn{\'e}odo}\ \emph {et~al.}(2008)\citenamefont
  {Arn{\'e}odo} \emph {et~al.}}]{arneodo}%
  \BibitemOpen
  \bibfield  {author} {\bibinfo {author} {\bibfnamefont {A.}~\bibnamefont
  {Arn{\'e}odo}} \emph {et~al.},\ }\bibfield  {title} {\enquote {\bibinfo
  {title} {Universal intermittent properties of particle trajectories in highly
  turbulent flows},}\ }\href@noop {} {\bibfield  {journal} {\bibinfo  {journal}
  {Phys.~Rev.~Lett.}\ }\textbf {\bibinfo {volume} {100}},\ \bibinfo {pages}
  {254504} (\bibinfo {year} {2008})}\BibitemShut {NoStop}%
\bibitem [{\citenamefont {Lanotte}\ \emph {et~al.}(2013)\citenamefont
  {Lanotte}, \citenamefont {Biferale}, \citenamefont {Boffetta},\ and\
  \citenamefont {Toschi}}]{lanotte}%
  \BibitemOpen
  \bibfield  {author} {\bibinfo {author} {\bibfnamefont {A.S.}\ \bibnamefont
  {Lanotte}}, \bibinfo {author} {\bibfnamefont {L.}~\bibnamefont {Biferale}},
  \bibinfo {author} {\bibfnamefont {G.}~\bibnamefont {Boffetta}}, \ and\
  \bibinfo {author} {\bibfnamefont {F.}~\bibnamefont {Toschi}},\ }\bibfield
  {title} {\enquote {\bibinfo {title} {A new assessment of the second-order
  moment of lagrangian velocity increments in turbulence},}\ }\href@noop {}
  {\bibfield  {journal} {\bibinfo  {journal} {J. Turb.}\ }\textbf {\bibinfo
  {volume} {14}},\ \bibinfo {pages} {34} (\bibinfo {year} {2013})}\BibitemShut
  {NoStop}%
\bibitem [{\citenamefont {Buaria}\ and\ \citenamefont {Pumir}(2022)}]{BP2022}%
  \BibitemOpen
  \bibfield  {author} {\bibinfo {author} {\bibfnamefont {D.}~\bibnamefont
  {Buaria}}\ and\ \bibinfo {author} {\bibfnamefont {A.}~\bibnamefont {Pumir}},\
  }\bibfield  {title} {\enquote {\bibinfo {title} {Vorticity-strain rate
  dynamics and the smallest scales of turbulence},}\ }\href@noop {} {\bibfield
  {journal} {\bibinfo  {journal} {Phys.~Rev.~Lett.}\ }\textbf {\bibinfo
  {volume} {128}},\ \bibinfo {pages} {094501} (\bibinfo {year}
  {2022})}\BibitemShut {NoStop}%
\bibitem [{Note4()}]{Note4}%
  \BibitemOpen
  \bibinfo {note} {The forcing term has negligible contribution to acceleration
  moments. This is also reaffirmed by collapse of data in Fig.~\ref {fig:acc2}
  from various sources that use different forcings}\BibitemShut {NoStop}%
\bibitem [{\citenamefont {Monin}\ and\ \citenamefont {Yaglom}(1975)}]{MY.II}%
  \BibitemOpen
  \bibfield  {author} {\bibinfo {author} {\bibfnamefont {A.~S.}\ \bibnamefont
  {Monin}}\ and\ \bibinfo {author} {\bibfnamefont {A.~M.}\ \bibnamefont
  {Yaglom}},\ }\href@noop {} {\emph {\bibinfo {title} {Statistical Fluid
  Mechanics}}},\ Vol.~\bibinfo {volume} {2}\ (\bibinfo  {publisher} {MIT
  Press},\ \bibinfo {year} {1975})\BibitemShut {NoStop}%
\bibitem [{\citenamefont {Hill}\ and\ \citenamefont
  {Wilczak}(1995)}]{hill1995}%
  \BibitemOpen
  \bibfield  {author} {\bibinfo {author} {\bibfnamefont {R.~J.}\ \bibnamefont
  {Hill}}\ and\ \bibinfo {author} {\bibfnamefont {J.~M.}\ \bibnamefont
  {Wilczak}},\ }\bibfield  {title} {\enquote {\bibinfo {title} {Pressure
  structure functions and spectra for locally isotropic turbulence},}\
  }\href@noop {} {\bibfield  {journal} {\bibinfo  {journal} {J.~Fluid Mech.}\
  }\textbf {\bibinfo {volume} {296}},\ \bibinfo {pages} {247--269} (\bibinfo
  {year} {1995})}\BibitemShut {NoStop}%
\bibitem [{\citenamefont {Antonia}\ \emph {et~al.}(1981)\citenamefont
  {Antonia}, \citenamefont {Chambers},\ and\ \citenamefont
  {Satyaprakash}}]{antonia1981}%
  \BibitemOpen
  \bibfield  {author} {\bibinfo {author} {\bibfnamefont {R.~A.}\ \bibnamefont
  {Antonia}}, \bibinfo {author} {\bibfnamefont {A.~J.}\ \bibnamefont
  {Chambers}}, \ and\ \bibinfo {author} {\bibfnamefont {B.~R.}\ \bibnamefont
  {Satyaprakash}},\ }\bibfield  {title} {\enquote {\bibinfo {title} {Reynolds
  number dependence of high-order moments of the streamwise turbulent velocity
  derivative},}\ }\href@noop {} {\bibfield  {journal} {\bibinfo  {journal}
  {Bound.-Layer Meteorol.}\ }\textbf {\bibinfo {volume} {21}},\ \bibinfo
  {pages} {159} (\bibinfo {year} {1981})}\BibitemShut {NoStop}%
\end{thebibliography}

%

\end{document}


\title{
Supplementary Material for \\
Scaling of acceleration statistics
in high Reynolds number turbulence
}

\author{Dhawal Buaria }
\affiliation{Tandon School of Engineering, New York University, New York, NY 11201, USA}
\affiliation{Max Planck Institute for Dynamics and Self-Organization, 37077 G\"ottingen, Germany}
\author{Katepalli R. Sreenivasan}
\affiliation{Tandon School of Engineering, New York University, New York, NY 11201, USA}
\affiliation{Department of Physics and the Courant Institute of Mathematical Sciences, New York University, New York, NY 10012, USA}

\maketitle


\section{Derivation for scaling of acceleration moments} 

The underlying idea behind all approaches is that
acceleration $a$ can be dimensionally written in 
terms of dissipation rate $\epsilon$ and viscosity $\nu$ as
\begin{align}
a \sim \epsilon^{3/4} \nu^{-1/4}.
\end{align}
Thus, the moments of acceleration are obtained as
\begin{align}
\langle a^p \rangle  \sim 
\langle \epsilon^{3p/4} \rangle \nu^{-p/4}.
\end{align}
Normalizing both side by Kolmogorov scales, we have
\begin{align}
\langle a^p \rangle /a_{\rm K}^p  \sim 
\langle \epsilon^{3p/4} \rangle / \langle \epsilon \rangle^{3p/4} 
\end{align}
where $a_{\rm K} = \langle \epsilon\rangle^{3/4} \nu^{-1/4}$.
Thus, our goal is to obtain the normalized moments
of dissipation. 
We present here the results from various intermittency models,
providing extra details compared to the main text.

\subsection{Multifractal approach based on velocity increments}

Within the multifractal framework, the velocity increment $\delta u_r$ over 
scale $r$ is assumed to be H\"older continuous:
\begin{align}
\delta u_r/u' \sim (r/L)^h
\label{eq:du}
\end{align}
where $h$ is the local H\"older exponent 
and $D(h)$ is the corresponding multifractal spectrum. 
Within this paradigm, a scale-dependent 
dissipation $\epsilon_r$ can be
defined as
\begin{align}
\epsilon_r \sim (\delta u_r)^3/r \ , 
\end{align}
which can be rewritten as
\begin{align}
\epsilon_r \sim 
\langle \epsilon \rangle \left(\frac{r}{L} \right)^{3h-1} \ ,
\label{eq:eps_du}
\end{align}
where we have used 
$\langle \epsilon \rangle \sim {u'^3}/{L}$ 
based on dissipation anomaly. 
We can obtain the moments of $\epsilon_r$ \cite{Frisch95} as:
\begin{align}
\langle \epsilon_r^q \rangle \sim 
\langle \epsilon \rangle^q
\int_h \left (\frac{r}{L} \right)^{q(3h-1)} 
\left( \frac{r}{L} \right)^{3-D(h)} dh \ .
\end{align}

Now $\epsilon_r$ reduces to true dissipation when $r$ corresponds 
to the viscous cutoff, often defined by the condition
\begin{align}
\delta u_r r /\nu \simeq 1 \ ,
\end{align}
which can be rewritten as
\begin{align}
(r/L) \simeq Re^{-1/(1+h)} \ ,
\end{align}
where $Re = u'L/\nu$.
Using the steepest gradient argument, 
and $Re \sim R_\lambda^2$, the scaling of the dissipation 
moments at large $Re$ can be obtained as
\begin{align}
\langle \epsilon^q \rangle 
/\langle \epsilon \rangle^q \sim  R_\lambda^{\tau_q} \ \ ,
\ \ \ \ \ \ \text{with} \ \ 
\tau_q = \sup_h \frac{2[D(h) - 3  + q(1-3h)]}{1+h}.
\end{align}
Thus, it readily follows that
\begin{align}
\langle a^p \rangle/a_{\rm K}^p 
\sim R_\lambda^{\zeta_p}  \ \ ,
\ \ \ \ \ \ \text{with} \ \ 
\zeta_p = \tau_{3p/4} \ .
\end{align}

The exponents can be obtained for any standard $D(h)$. 
A well known approximation is given by 
the She-Leveque model \cite{SL94}:
\begin{align}
D(h) =  \frac{3(h-h^*)}{\log \gamma} 
\left[ \log \left( \frac{3(h-h^*)}{d^* \log \gamma} \right) - 1 \right]
+ 3 - d^*
\label{eq:shelev}
\end{align}
where $d^* = (1-3h^*)/(1-\gamma)$,
$h^* = 1/9$ and $\gamma = 2/3$.

\vspace{0.1in}
\paragraph*{Longitudinal versus transverse directions}:
In principle, the multifractal
model does not
differentiate between longitudinal
and tranverse directions, 
i.e., it predicts that velocity increments
and gradients in both longitudinal and
transverse directions scale similarly.
However, it is now well established that this is not
the case,with tranverse increments/gradients being 
somewhat more intermittent \cite{Ishihara07, arneodo, BPBY2019, BP2022}. 
Based on this, an ad-hoc modification
to $D(h)$ in Eq.~\eqref{eq:shelev}
was proposed in ref.~\cite{arneodo},
with $h^*=1/9$ but $\gamma=1/2$, 
to empirically fit the scaling
exponents of transverse structure functions.
Thereafter, this result was later utilized
to predict the scaling of acceleration
variance in \cite{lanotte}.
However, we note that this extension 
is unjustified, since this modified
$D(h)$ (originally intended to 
describe transverse structure functions) 
already fails to describe the scaling 
of transverse velocity gradients.
For instance, this modified
$D(h)$ predicts 
the following for enstrophy:
$\langle \Omega^{3/2} \rangle /\langle \Omega \rangle^{3/2} 
\sim R_\lambda^{0.23}$ (as opposed to the $0.14$
predicted for scaling dissipation).
However, this prediction is at clear odds with 
DNS data in Fig.~3a, which shows that
both dissipation and enstrophy moments
are close to the $0.14$ prediction.
Essentially, this ad-hoc modification to $D(h)$  
grossly overpredicts the scaling
of transverse velocity gradients
and hence cannot be utilized to describe the scaling
of acceleration also.

\subsection{Multifractal approach based on dissipation}

The multifractal approach based on dissipation
directly deals with the dissipation field,
with the starting point being \cite{MS91}
\begin{align}
\epsilon_r / \langle \epsilon \rangle \sim (r/L)^{\alpha-1} \ ,
\label{eq:epsr_alpha}
\end{align}
where $\alpha$ is the local H\"older exponent,
with the multifractal spectrum
given by some $F(\alpha)$. 
In the literature \cite{MS91},
the 1D spectrum $f(\alpha)$ is more
commonly used, which is simply given by
$f(\alpha) = F(\alpha)-2$.
As is evident, this expression is essentially equivalent to
that in Eq.~\eqref{eq:eps_du}, with $\alpha = 3h$. 
It also follows that the moments
of $\epsilon_r$ are given as 
\begin{align}
\langle \epsilon_r^q \rangle \sim 
\langle \epsilon \rangle^q
\int_\alpha \left (\frac{r}{L} \right)^{q(\alpha-1)} 
\left( \frac{r}{L} \right)^{3-F(\alpha)} d\alpha \ .
\end{align}

Now, the viscous cutoff 
at which $\epsilon_r$ reduces
to true dissipation 
is defined by the scale
\begin{align}
r = (\nu^3/\epsilon_r)^{1/4} \ ,
\end{align}
which, by using Eq.~\eqref{eq:epsr_alpha}, 
leads to
\begin{align}
r/L \simeq Re^{-3/(3+\alpha)} \ .
\end{align}
Here, we have also used
$\langle \epsilon \rangle \sim u'^3/L$
and $Re = u'L/\nu $.
Using the steepest gradient argument, 
and $Re \sim R_\lambda^2$, the scaling of the dissipation 
moments at large $Re$ can be obtained as
\begin{align}
\langle \epsilon^q \rangle 
/\langle \epsilon \rangle^q \sim  R_\lambda^{\tau_q} \ \ ,
\ \ \ \ \ \ \text{with} \ \ 
\tau_q = \sup_\alpha \frac{6[F(\alpha) - 3  + q(1-\alpha)]}{3+\alpha} \ .
\end{align}

In the literature, the result is often
stated in terms of the Renyi dimension $D_q$,
which is the Legendre transform of
$F(\alpha)$.
Stated in terms of $D_q$ the previous result
can be rewritten as \cite{borgas93}
\begin{align}
\tau_q = 6(\tilde{q} - q) \ ,
\end{align}
where the unique function $\tilde{q}(q)$ is 
determined from 
\begin{align}
4 (\tilde{q} - q) = (\tilde{q} - 1) (1 - D_{\tilde{q}}) \ .
\end{align}
An approximation for $D_q$ is provided by the p-model \cite{MS87}:
\begin{align}
D_q = \frac{1}{1-q} \log_2 (p_1^q + p_2^q) \ ,
\end{align}
for $p_1=0.7$ and $p_2 = 1-p_1$.

\subsection{Kolmogorov's lognormal model}

From the Kolmogorov (1962) lognormal model \cite{K62},
it is well-known that 
\begin{align}
\langle \epsilon^q \rangle 
/\langle \epsilon \rangle^q \sim  R_\lambda^{\tau_q} \ \ ,
\ \ \ \ \ \ \text{with} \ \ 
\tau_q = \frac{3}{4} \mu q(q-1) \ ,  
\end{align}
where $\mu$ is the intermittency exponent,
with standard estimates suggesting $\mu=0.25$ \cite{sreeni93}. 
This result also corresponds
to $D_q = 1 - \mu q /2$.

\section{Integration of fourth order structure function}

\begin{figure}[h]
\centering
\includegraphics[width=0.5\textwidth]{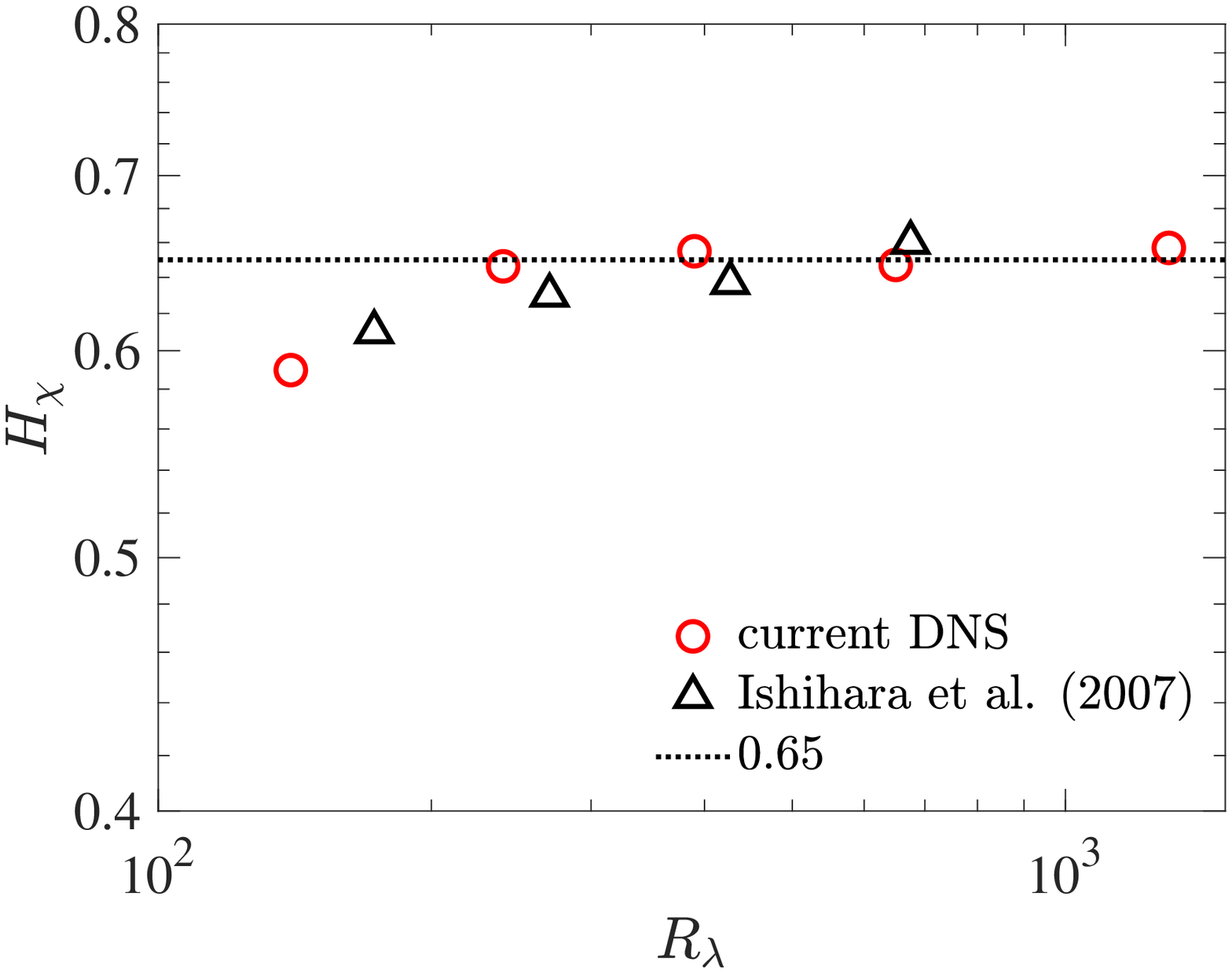} 
\caption{
Plot of $H_\chi$ vs. $\re$.
}
\label{fig:hchi}
\end{figure}

As noted in the main text,
acceleration variance can be
obtained by piecewise integration
of the fourth-order structure
function $D_{1111}(r)$:
\begin{align}
a_0 = \frac{\langle a^2 \rangle}{ (\langle \epsilon \rangle^{3/2} \nu^{-1/2}) }
= \frac{4 H_\chi}{3}  \int_r  \left( \frac{r}{\eta_K} \right)^{-3} 
\frac{D_{1111} (r)}{u_K^4} d \left( \frac{r}{\eta_K}\right) \ ,
\label{eq:acc_d4}
\end{align}
where $H_\chi$ is defined
by the above relation and is assumed 
to be constant
at sufficiently high $\re$.
This is confirmed by previous
\cite{Ishihara07} and current
DNS data, both plotted in
Fig.~\ref{fig:hchi}, 
showing
$H_\chi \approx 0.65$  for $\re\gtrsim200$.

The right hand side in Eq.~\eqref{eq:acc_d4} can be
obtained via piecewise integration,
utilizing:
\begin{align}
\frac{D_{1111}(r)}{u_K^4} =
  \begin{cases}
\frac{F}{225} \left( \frac{r}{\eta_K} \right)^4 
\ \ \ \ \ \    &r < \ell \ , \\
C_4 \left( \frac{r}{\eta_K} \right)^{\xi_4} 
\ \ \ \ \ \    &\ell < r < L \ , \\
C 
\ \ \ \ \ \ \  &r > L \ , 
\end{cases}
\label{eq:d4}
\end{align}
where $F$ is the flatness of $\partial u/\partial x$, 
$\xi_4$ is the corresponding inertial-range
exponent. Note that both $C_4$ and $C$ are functions of $\re$. 
The scale $\ell$ corresponds to the crossover
between dissipation and inertial range, and is
obtained by invoking continuity of $D_{1111}(r)$
at $r=\ell$, i.e.,
$(F/225) ( {\ell}/{\eta_K})^4 =
C_4 ( {\ell}/{\eta_K} )^{\xi_4}$,
which leads to
\begin{align} 
( {\ell}/{\eta_K} )^{4-\xi_4} = 225 \ C_4/F \ .
\label{eq:ellf}
\end{align}

To simplify notation, we momentarily take
$r/\eta_K \to r$,
$\ell/\eta_K \to \ell $, $L/\eta_K \to L$ and
evaluate the integral in Eq.~\eqref{eq:acc_d4} as
\begin{align} 
a_0 &= \int_0^{\ell} \frac{F}{15^2} r  dr
+ \int_{\ell}^{L} C_4 r^{\xi_4-3}  dr
+ \int_{L}^{\infty} C r^{-3}  dr \\
&= \frac{F}{15^2} \frac{\ell^2}{2}
+ \frac{C_4}{\xi_4-3+1} \left( L^{\xi_4-3+1} - \ell^{\xi_4-3+1} \right)
+ \frac{C}{-3+1} \left(0 - L^{-3+1} \right) \\
&= \frac{F}{15^2} \frac{\ell^2}{2}
+ \frac{C_4}{2-\xi_4} \ell^{\xi_4-2} 
- \frac{C_4}{2-\xi_4} L^{\xi_4-2} 
+ \frac{C}{2 L^2}.  
\end{align}
As $\xi_4<2$, we can ignore
the last two terms with $L$, since
they will both decrease with increasing $\re$
(note $C \sim \re^2$ and $L \sim \re^{3/2}$ \cite{Frisch95}).
Thus, we get
\begin{align} 
a_0
= \frac{F}{15^2} \frac{\ell^2}{2}
+ \frac{C_4}{2-q} \ell^{\xi_4-2}.
\label{eq:chi2}
\end{align}
Note that we have $F \ell^4 \sim C_4 \ell^{\xi_4}$ from Eq.~\eqref{eq:ellf},
which implies
$F \ell^2 \sim C_4 \ell^{\xi_4-2}$.
Thus, both terms in Eq.~\eqref{eq:chi2}
have the same scaling,
and we can simply write
\begin{align} 
a_0 \simeq F (\ell/\eta_K)^2,
\label{eq:a0f}
\end{align}
where we have replaced
$\ell \to \ell/\eta_K$.

The $\re$ scaling of $F$ and $C_4$ 
are now taken to be
\begin{align}
F \sim \re^\alpha \ , \ \ \ \ \  C_4 \sim \re^\beta 
\end{align}
which, upon substitution in Eq.~\eqref{eq:ellf}, 
also gives
\begin{align} 
 \frac{\ell}{\eta_K} \sim   \re^{(\beta - \alpha)/(4-\xi_4)}.
\end{align}
Finally, substituting these relations in Eq.~\eqref{eq:a0f}, we get
\begin{align} 
a_0 \sim   \re^{(2\alpha - \alpha q + 2\beta)/(4-\xi_4)},
\label{eq:a0f2}
\end{align}
which is reported in the main text.
Using $\alpha\approx0.387$, $\beta\approx0.2$
and $\xi_4 \approx 1.30$, we get
\begin{align} 
a_0 \sim   \re^{0.25} \ .
\end{align}

\section{Updated interpolation formula for $a_0$ vs. $\re$}

As evident from Fig.~1 (and Fig.~3a) of the main text,
the result $a_0 \sim   \re^{0.25}$ describes
the data excellently for $\re \gtrsim 200$.
To approximately model the data for lower $\re$, it might be
worthwhile to consider an interpolation
formula similar to Eq.~(3) of the main text,
with $\chi=0.25$. A least-square fit
gives $c_1\approx0.89$ and $c_2\approx40$.
The fit is shown in the figure below.
We note that this result is only meant to be
an empirical approximation. In principle,
other interpolation formulae can also be devised
to obtain a more robust fit at low $\re$.

\begin{figure}[h]
\centering
\includegraphics[width=0.45\textwidth]{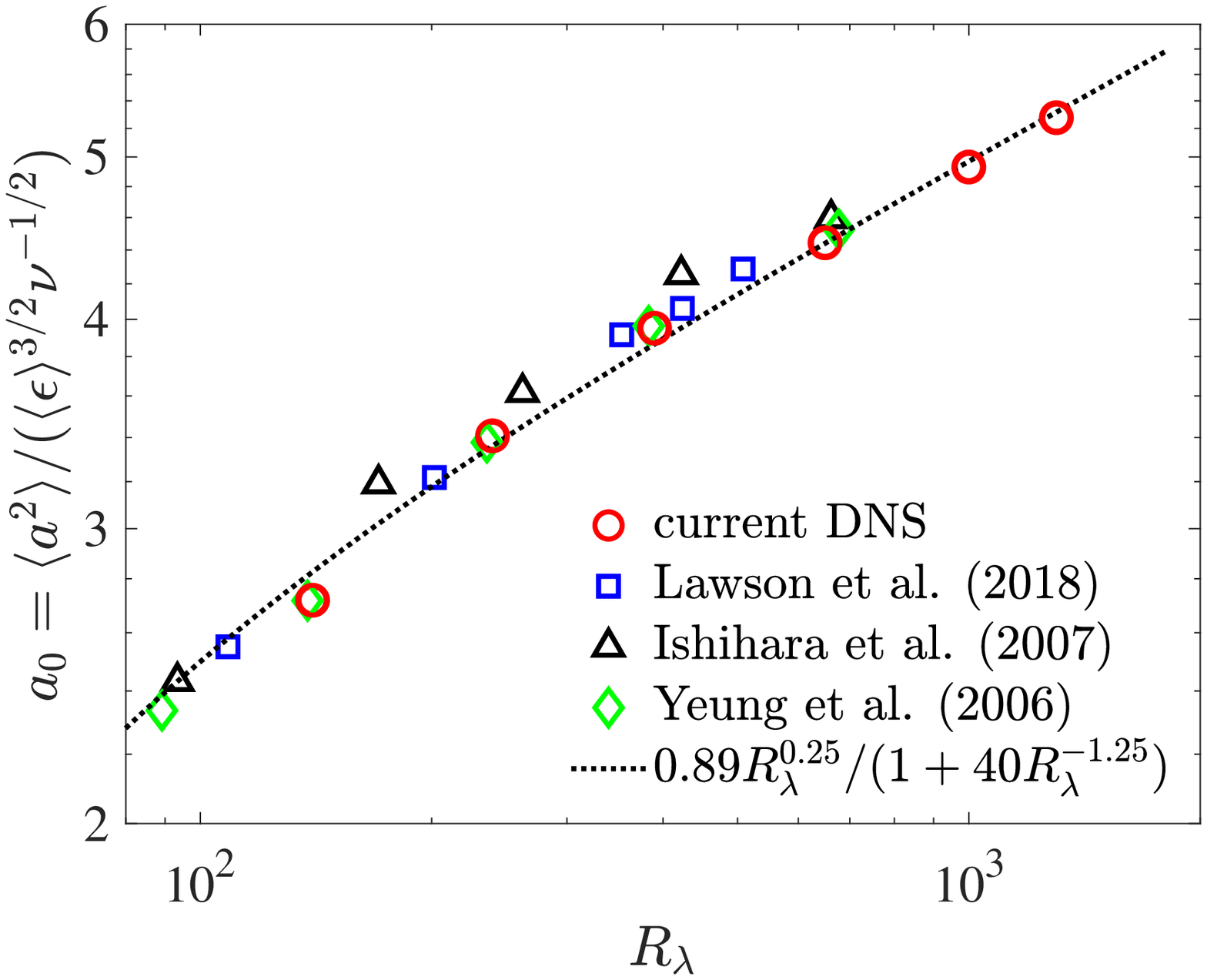} 
\caption{
Plot of $a_0$ vs. $\re$ showing how the updated 
interpolation formula, corresponding to Eq.~(3) of main text,
fits the available data.
}
\label{fig:accfit}
\end{figure}

\section{Scaling of structure function proportionality constants}

The inertial-range scaling of $p$-th order structure functions 
can be written as:
\begin{align}
\frac{\langle (\delta u_r)^p \rangle }{ u_K^p}  = C_p \left( \frac{r}{\eta_K} \right)^{\zeta_p} \ ,
\label{eq:strfn_p}
\end{align}
where $C_p$ are the proportionality constants and $\zeta_p$ is the inertial range.
As per K41, $\zeta_p = p/3$ and $C_p$ are universal constants independent of $\re$.
However, due to intermittency $\zeta_p$ is a non-trivial function of $p$ and
$C_p$ depends on $\re$ (except for $p=3$). 

To derive the $\re$-scaling of $C_p$, we start again from Eq.~\eqref{eq:du}.
Standard multifractal arguments \cite{Frisch95} give:
\begin{align}
\langle (\delta u_r)^p \rangle \sim {u^\prime}^p  \left(\frac{r}{L} \right)^{\zeta_p}  \ ,     
\ \ \ \ \ \ \text{with} \ \ 
\zeta_p = \inf_h  \ [ph + 3  - D(h)] \ .
\end{align}
Normalizing the above equation by Kolmogorov variables gives:
\begin{align}
\frac{\langle (\delta u_r)^p \rangle}{u_K^p} \sim 
\left( \frac{ {u^\prime}}{u_K} \right)^p    
\left(\frac{\eta_K}{L} \right)^{\zeta_p}
\left(\frac{r}{\eta_K} \right)^{\zeta_p} \ ,     
\end{align}
Thereafter, using $u^\prime/u_K \sim \re^{1/2}$ and $L/\eta_K \sim \re^{3/2}$ 
\cite{Frisch95} leads to:
\begin{align}
\frac{\langle (\delta u_r)^p \rangle}{u_K^p} \sim 
R_\lambda^{p/2 - 3\zeta_p/2} \   \left(\frac{r}{\eta_K} \right)^{\zeta_p}  \ .
\end{align}
Comparing with Eq.~\eqref{eq:strfn_p} gives:
\begin{align}
C_p \sim  R_\lambda^{(p - 3\zeta_p)/2}  \ .
\end{align}
For $p=4$, we have $C_4 \sim \re^\beta$. Thus, the multifractal prediction
gives $\beta = (4-3\zeta_4)/2$. 
For K62 log-normal, we have $\zeta_p = 4/3 - 2\mu/9$, giving $\beta=\mu/3$.
Both multifractals and K62 log-normal predict $\zeta_4\approx1.28$ \cite{Frisch95},
leading to $\beta \approx 0.08$, which is substantially lower than 
the $\beta\approx0.2$ suggested by DNS results, as reported in the main text.


%